\documentclass{IEEEoj-data}
\usepackage{cite}
\usepackage{amsmath,amssymb,amsfonts}
\usepackage{graphicx}%
\usepackage{textcomp}
\usepackage{booktabs}
\usepackage{array}%
\usepackage{siunitx}
\usepackage{orcidlink}
\usepackage{url}%
\usepackage[caption=false,font=footnotesize]{subfig}
\makeatletter
\renewcommand*{\p@subfigure}[1]{\thefigure(#1)}
\makeatother
\usepackage{hyperref}
\hypersetup{pdftitle={Descriptor: UWB Anchor Self-Calibration Dataset (UWB-ASC)},
  pdfauthor={Paul Schwarzbach}}
\usepackage[capitalize]{cleveref}
\graphicspath{{figures/}}

\makeatletter
\def\endthebibliography{\def\@noitemerr{\@latex@warning{Empty bibliography}}\endlist}
\makeatother

\newcommand{\WEnodeAnchorsQsSetA}{11}
\newcommand{\WEnodeAnchorsStaticSetA}{12}
\newcommand{\WEnodeAnchorsSetB}{10}

\newcommand{\DFrangingNodesSetA}{12}

\newcommand{\WEresAaMedSetB}{0.27}
\newcommand{\WEresMaMedSetB}{0.24}

\newcommand{\DFresObstructedSpanSetB}{4.0--5.8}
\newcommand{\WEresAaMedSetA}{0.28}

\newcommand{\DFresAaBiasedPairsSetA}{6}

\newcommand{\WEgtViconRateSpan}{119.7--120.1}
\newcommand{\DFviconMeanRateSpan}{23.5--24.1}
\newcommand{\WEgtClockSigma}{0.1}

\newcommand{\DFclockOffsetSpanSetB}{6.85--6.89}

\newcommand{\DFinvalidPctSetB}{0.066}

\newcommand{\DFinvalidPctSetA}{0.093}

\newcommand{\DFquasiStaticSetA}{20}

\newcommand{\DFlabelPairsSetA}{66}

\newcommand{\DFlabelMixedSetA}{1}
\newcommand{\DFlabelPairsSetB}{55}

\newcommand{\DFlabelMixedSetB}{3}

\newcommand{\WEsynSigmaHw}{0.33}
\newcommand{\WEsynPoutPct}{5}
\newcommand{\WEsynDraws}{20}
\newcommand{\WEsynEpochs}{200}
\newcommand{\DFsynSessions}{81}
\newcommand{\DFsynConditions}{5}
\newcommand{\DFsynBias}{0.27}
\newcommand{\DFsynAnchors}{10}
\newcommand{\DFsynPairs}{45}

\newcommand{\DFsynMpMean}{0.8}
\newcommand{\DFsynMpSigma}{1.07}
\newcommand{\DFsynDmax}{100}
\newcommand{\DFextentLongSetA}{36}
\newcommand{\DFextentShortSetA}{27}
\newcommand{\DFheightSpanCmSetA}{10}
\newcommand{\DFextentLongSetB}{19}
\newcommand{\DFextentShortSetB}{12}
\newcommand{\DFheightSpanCmSetB}{4}
\newcommand{\DFsigmaSurveyCm}{1}

\newcommand{\DFyearSetB}{2025}
\newcommand{\DFmatchTolS}{0.06}
\newcommand{\DFmaxGapS}{0.5}

\newcommand{\DFresAaBiasThresholdM}{5}

\usepackage{etoolbox}
\makeatletter
\def\ieeelogo{}
\patchcmd{\ps@headings}{VOLUME\ 00\@jvol,\ 2024\@pubyear}{}{}{\errmessage{arXiv build: the footer of headings was not patched}}
\patchcmd{\ps@headings}{VOLUME\ 00\@jvol,\ 2024\@pubyear}{}{}{\errmessage{arXiv build: the footer of headings was not patched}}
\patchcmd{\ps@plain}{VOLUME\ 00\@jvol,\ 2024\@pubyear}{}{}{\errmessage{arXiv build: the footer of plain was not patched}}
\patchcmd{\ps@plain}{VOLUME\ 00\@jvol,\ 2024\@pubyear}{}{}{\errmessage{arXiv build: the footer of plain was not patched}}
\ps@headings
\makeatother
\makeatletter
\AddToHookNext{shipout/foreground}{\put(\strip@pt\dimexpr306pt-\hoffset\relax,\strip@pt\dimexpr\voffset-52pt\relax){\makebox[0pt][c]{\parbox[b]{430pt}{\centering\fontsize{7.5}{9}\selectfont\itshape This work has been submitted to the IEEE for possible publication. Copyright may be transferred without notice, after which this version may no longer be accessible.}}}}
\makeatother
\begin{document}

\title{{\color{black}\textbf{Descriptor:}} \textit{UWB Anchor Self-Calibration Dataset} (UWB-ASC)}

\author{Paul Schwarzbach\,\orcidlink{0000-0002-1091-782X}\authorrefmark{1}}
\affil{Chair of Transport Systems Information Technology, TUD Dresden University of Technology, Dresden, Germany}
\corresp{CORRESPONDING AUTHOR: Paul Schwarzbach (e-mail: paul.schwarzbach@tu-dresden.de).}
\markboth{DESCRIPTOR: UWB ANCHOR SELF-CALIBRATION DATASET (UWB-ASC)}{SCHWARZBACH}

\begin{abstract}
Anchor self-calibration estimates the positions of fixed nodes of an indoor
radio positioning or localization system from observations among the nodes and
to mobile devices. It replaces the survey of every anchor, which is costly and
goes stale when anchors move. The UWB Anchor Self-Calibration Dataset
(UWB-ASC) is a testing ground for such methods, recorded in one industrial hall
with ultra-wideband (UWB) two-way ranging. It has three parts. The first is two
measured deployments. In the line-of-sight (LOS) deployment,
\WEnodeAnchorsSetB{} anchors range with a walking tag tracked by motion capture.
In the non-line-of-sight (NLOS) deployment, \DFrangingNodesSetA{} nodes range
among obstructions, and one of them roves over surveyed markers in
\DFquasiStaticSetA{} quasi-static sessions. Both deployments also hold static
sessions, and a total station surveyed every fixed node. The second part is a
seeded range simulator that draws synthetic ranges on the surveyed geometry with
adjustable error components, from noise-free ranges to NLOS errors, for
benchmarks at controlled error levels. Its \DFsynSessions{} frozen draws in
\DFsynConditions{} conditions serve as the reference. The third part is a set
of readers that convert the ranges among the static nodes of two public UWB
datasets, TWR-CLOUD-Bern and selfloc, into the same format without
redistribution. All parts share open Parquet and YAML files, in which
topology views select the ranges for each class of calibration method. Raw
recordings with rebuild scripts, quality reports, and field labels of the line
of sight complete the dataset.
\\
{\textcolor{ieeedata}{\abstractheadfont\bfseries{IEEE SOCIETIES/COUNCILS}}}\quad Sensors Council (SENC); Signal Processing Society (SPS)
\\
{\textcolor{ieeedata}{\abstractheadfont\bfseries{DATA DOI/PID}}}\quad \href{https://doi.org/10.5281/zenodo.24055896}{10.5281/zenodo.24055896}
\\
{\textcolor{ieeedata}{\abstractheadfont\bfseries{DATA TYPE/LOCATION}}}\quad UWB two-way ranges with surveyed and motion-capture ground truth (Parquet, YAML); industrial hall, Torgau, Germany
\end{abstract}


\maketitle

\section*{BACKGROUND}

\subsection*{Anchor Self-Calibration}
Radio positioning systems locate mobile devices from observations between them
and fixed infrastructure nodes, such as ranges, angles, or signal strengths, and
need the positions of these nodes. In ultra-wideband (UWB) systems with two-way
ranging, the observations are ranges between pairs of modules, and the fixed
nodes are called anchors~\cite{ridolfi_self-calibration_2021}.

The anchor positions are usually surveyed with an optical instrument like a
total station after installation. The survey binds trained staff and equipment
to every anchor and mainly drives the deployment cost of UWB
systems~\cite{ridolfi_self-calibration_2021}. It is slow and prone to errors
in large systems~\cite{herbruggen_multihop_2023}, and it must be repeated when
anchors are moved or added~\cite{krapez_anchor_2020}. According to an industry
analysis, the cost of deploying and installing a real-time locating system can
hold back its adoption more than the price of its
infrastructure~\cite{zignani_5g_2021}.

Anchor self-calibration addresses this cost. It estimates the anchor positions
from ranges that the anchors measure among themselves, from ranges to mobile
nodes, or from both~\cite{ridolfi_self-calibration_2021}. The literature
describes this one operation under many names, among them auto-positioning,
self-localization, and self-positioning~\cite{schwarzbach_terminology_2026}.
Methods differ in the measurement topology they use, from a full mesh of anchor
ranges over meshes whose anchors reach one another over multiple hops to the
ranges of a single moving tag, and in the reference information they assume,
such as a few surveyed anchors~\cite{schwarzbach_terminology_2026}. The author's
work includes a Bayesian auto-positioning
approach~\cite{jung_auto-positioning_2022}, grid-based uncertainty propagation
for collaborative self-calibration~\cite{jung_uncertainty_propagation_2025},
and a statistical model of UWB range errors that simulates
ranges~\cite{schwarzbach_statistical_2022}.

\subsection*{Related Datasets}
A testing ground for anchor self-calibration needs observations of each
topology from one node set, surveyed truth for every node, static and moving
nodes, and propagation conditions that bias the ranges as real sites do. Open
data of this kind are scarce. In ORDIP, Anagnostopoulos et~al.\ catalogued the
open data of indoor positioning and found UWB datasets few and mostly from
office environments~\cite{anagnostopoulos_ordip_2025}. Their review of the
papers of recent IPIN conferences found open UWB data rarely used and
industrial environments severely
underrepresented~\cite{anagnostopoulos_comprehensive_2025}. Both studies
recommend best practices to the providers of such data, and UWB-ASC follows
them, as RECORDS AND STORAGE shows.

Several public UWB datasets cover part of these needs.
TWR-CLOUD-Bern~\cite{xenakis_cooperative_2026,xenakis_twr-cloud-bern_2026}
provides two-way ranges among static DW3000 nodes, with truth from a laser
scan, and the selfloc
testbeds~\cite{corbalan_self-localization_2023,corbalan_selfloc_2023} provide
ranges among DW1000 anchors that reach one another over multiple hops, with
manually determined positions and no stated license.
UTIL~\cite{zhao_util_2024} records time-difference-of-arrival measurements of a
moving tag with motion-capture truth, and MILUV~\cite{shalaby_miluv_2026} and
the outdoor set of Lee et~al.~\cite{lee_outdoor_2025} record ranges between
tags and anchors, but none of the three holds ranges among anchors. The
AAU meshed set~\cite{jung_aau_meshed_2024,jung_modular_2024} combines ranges
among anchors with ranges to an unmanned aerial vehicle that motion capture
tracks, on the same module family as UWB-ASC, in a small flight arena with a
strongly three-dimensional layout. The HYMN
dataset~\cite{ammad_hymn_descriptor_2026} comes from the hall of UWB-ASC and
serves positioning with several radio technologies at surveyed reference
points. Of these, the AAU set comes
closest to the needs above, and UWB-ASC adds an industrial hall with static,
quasi-static, and walking sessions on one node set under LOS and NLOS
propagation.

\subsection*{Contribution and Reuse}
UWB-ASC is a testing ground for anchor self-calibration. Its three parts share
one format.
\begin{enumerate}
\item Two measured deployments in one hall. The line-of-sight (LOS)
  deployment, \texttt{set\_b} in the files, has a mostly unobstructed layout
  through which a tag was carried on walks. The non-line-of-sight (NLOS)
  deployment, \texttt{set\_a}, stands among obstructions, and a tag roves over
  surveyed markers. Both hold static sessions, and a total station
  surveyed every fixed node.
\item A seeded range simulator that draws synthetic ranges on the surveyed
  geometry with error components that a user sets, from noise-free ranges to
  NLOS errors. Frozen draws serve as its reference.
\item Readers that convert the ranges of TWR-CLOUD-Bern and selfloc into the
  format of UWB-ASC, so that anchor meshes of another chip generation and with
  multiple hops complement the measured deployments.
\end{enumerate}

The static sessions of the NLOS deployment served the author's study of
grid-based uncertainty propagation for collaborative
self-calibration~\cite{jung_uncertainty_propagation_2025}. The quasi-static
sessions of that deployment and all sessions of the LOS deployment are
published here for the first time, and the author's companion tutorial in
preparation uses them. The range simulator ports the statistical
range-error model~\cite{schwarzbach_statistical_2022} to seeded code.

The data support the calibration of anchor positions from anchor ranges,
mobile ranges, or both, under LOS and NLOS propagation. The
surveyed truth also supports studies of the range error itself, such as the
per-pair bias of two-way ranging or the detection of NLOS links against the
field labels, and views that select the ranges of each topology make the
topology a controlled variable of such studies. In calibration, the anchors
that fix the frame of an estimate are a controlled variable too. The range simulator sets the error level
of a test down to noise-free ranges, which separate the error of a method from
that of the data. INSIGHTS AND NOTES states what the testing ground lacks.

\section*{COLLECTION METHODS AND DESIGN}
Table~\ref{tab:parts} sets the three parts of UWB-ASC side by side. They share
one file format, so one loader reads every part and the same topology views
select their ranges. The Zenodo record holds the measured deployments and the
frozen synthetic draws, while the range simulator and the readers write further
sessions to a folder of the user's choice.

\begin{table}[!htb]
  \caption{Data of the testing ground with their topology views
    (Fig.~\ref{fig:views}). Anchor views hold anchor ranges only. Synthetic rows
    hold no measurement.}
  \label{tab:parts}
  \centering
  \footnotesize
  \setlength{\tabcolsep}{2pt}
  \begin{tabular}{@{}llll@{}}
    \toprule
    Data & Mobile node & Truth & Views \\
    \midrule
    \multicolumn{4}{@{}l}{In the Zenodo record} \\
    LOS deployment & walking, static & survey, Vicon & all four \\
    NLOS deployment & on markers & survey, markers & all four in \texttt{s01}--\texttt{s20} \\
    Synthetic, frozen & none & exact & anchor views \\
    \midrule
    \multicolumn{4}{@{}l}{Written on demand by the code of the record} \\
    Synthetic, new & none or tag & exact & all four with a tag \\
    TWR-CLOUD-Bern & none & as released & two anchor views \\
    selfloc & none & as released & two anchor views \\
    \bottomrule
  \end{tabular}
\end{table}

\begin{figure*}[!t]
  \centering
  \subfloat[\texttt{fully\_meshed}\label{fig:views_a}]{%
    \includegraphics{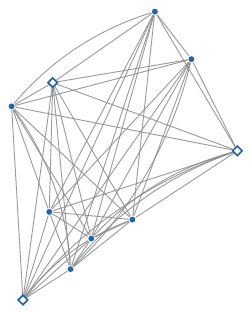}}\hfil
  \subfloat[\texttt{mobile\_assisted}\label{fig:views_b}]{%
    \includegraphics{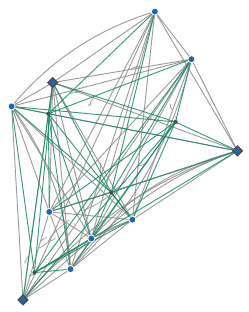}}\hfil
  \subfloat[\texttt{slam\_bipartite}\label{fig:views_c}]{%
    \includegraphics{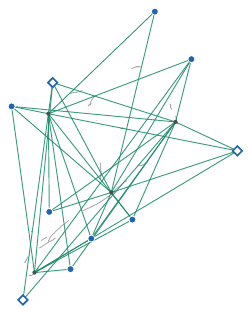}}\hfil
  \subfloat[\texttt{multi\_hop}\label{fig:views_d}]{%
    \includegraphics{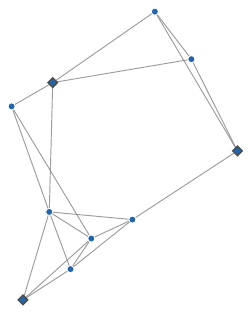}}\\[1pt]
  \includegraphics{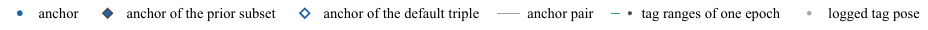}
  \caption{Topology views of the walk \texttt{set\_b/s01}, with the anchors that
    fix the datum. Anchor pairs whose straight line would run through a third
    anchor are drawn as arcs. Each panel lies in the survey frame of the LOS
    deployment.}
  \label{fig:views}
\end{figure*}

\subsection*{Common Format and Topology Views}
Every session of UWB-ASC, whether measured, synthetic, or converted, stores its
nodes with their roles and positional priors, its directed ranges, the truth of
its mobile node where one exists, and its metadata. A systematic review distinguishes
four measurement topologies~\cite{schwarzbach_terminology_2026}, and a topology
view selects the ranges and the priors of one of them from a session. It is a
short YAML mask that never adds data, so methods of every topology run on the
same session. Every campaign of the Zenodo record carries the four views, which
Fig.~\ref{fig:views} draws on one walk of the LOS deployment.
\begin{itemize}
\item \texttt{fully\_meshed} keeps the ranges among the anchors and none of
  their surveyed priors.
\item \texttt{mobile\_assisted} keeps the ranges among the anchors and those
  between the mobile node and the anchors, with the surveyed priors of a subset
  of three anchors across the layout (5BB6, 4A23, and 8B05,
  Fig.~\ref{fig:layout}), the same in both deployments.
\item \texttt{slam\_bipartite} keeps only the ranges between the mobile node and
  the anchors and no prior, as in simultaneous localization and mapping. With
  a standing tag it needs several sessions pooled.
\item \texttt{multi\_hop} thins the anchor mesh to the three nearest neighbors of
  each anchor by median range and keeps the priors of the same subset. In both
  deployments, the thinned mesh leaves some anchors free to turn or mirror
  against the rest, despite the priors.
\end{itemize}

\subsection*{Datum and Evaluation}
Ranges fix a layout only up to a rigid motion and a reflection, so an estimate
needs a datum that ties it to a frame~\cite{teunissen_zero_1985}. A view with
priors gives an absolute datum in the survey frame. Without
priors a method fixes a relative datum, and UWB-ASC follows the convention of
an earlier study~\cite{jung_uncertainty_propagation_2025}, in which a gauge
triple places its first anchor at the origin, the second on the positive
$x$-axis, and the third at positive $y$. The anchors of the prior subset form
the default triple, in the order 5BB6, 4A23, 8B05 (Fig.~\ref{fig:views}). In
the quasi-static sessions, where 8B05 is the roving tag, two anchors of the
triple remain, and they leave the reflection about the line through them open,
for both kinds of datum.

\begin{figure*}[!t]
  \centering
  \subfloat[LOS deployment (\texttt{set\_b})\label{fig:layout_a}]{%
    \includegraphics{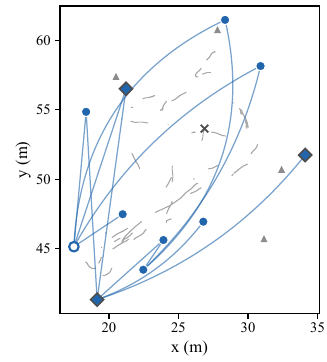}}\hfil
  \subfloat[NLOS deployment (\texttt{set\_a})\label{fig:layout_b}]{%
    \includegraphics{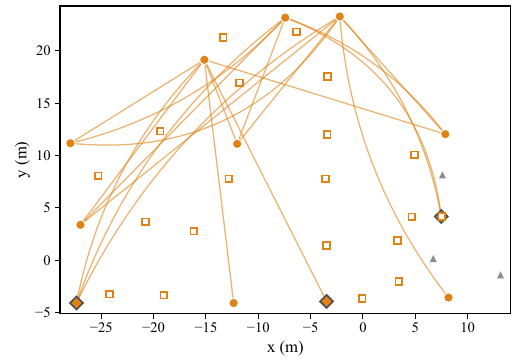}}\\[1pt]
  \includegraphics{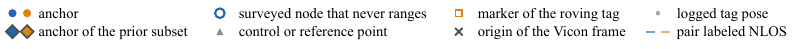}
  \caption{Layout of both deployments from the survey and the field labels,
    with the logged tag poses of the walks \texttt{set\_b/s01} and \texttt{s02}.
    Pairs labeled NLOS whose straight line would run through another mark are
    drawn as arcs. Each panel lies in the survey frame of its deployment.}
  \label{fig:layout}
\end{figure*}

The prior subset and the default triple are conventions of the testing ground.
Every fixed node is surveyed, so a study can fix the datum on other anchors,
with a mask or a triple of its own, and vary it as it varies the topology. The
choice matters for methods that build the layout outward from the first
anchors, since the errors of their ranges pass on to every anchor placed
later. The same study varied the triangle of the first three anchors on the
static sessions of the NLOS deployment, between triangles with all sides in LOS
and triangles with two sides in NLOS~\cite{jung_uncertainty_propagation_2025}.
By the field labels, the default triangle has all sides in LOS there, while in
the LOS deployment the labels mark both its sides at 5BB6, the origin of its
frame, as NLOS (Fig.~\ref{fig:layout_a}).

The evaluation convention aligns an estimate to the survey in the horizontal
plane by a rigid transform without scale and with a reflection
allowed~\cite{umeyama_least-squares_1991}. The alignment makes estimates of
every datum comparable, but also removes the error that the datum puts into the
frame. A study of the datum therefore compares an estimate that uses priors in
the survey frame without alignment, and one in the frame of a gauge triple against
the survey in that frame. The truth of the motion that a view keeps is for
evaluation only.

\subsection*{Hardware and Site}
Both deployments use the modules of a Qorvo MDEK1001 development
kit~\cite{qorvo_mdek1001}, which carry the DW1000 UWB chip and measure
distances by two-way ranging. The modules ran a modified firmware that is not
published, and neither its version nor the physical-layer settings (channel,
pulse repetition frequency, preamble, and data rate) were recorded. A logging
computer read the ranges through the decoder of the kit and wrote them to
backup logs, one record per ranging epoch, numbered by the frame counter
\texttt{seq}. Each record is a directed distance matrix, in which each row
lists the ranges that one node measured to the other nodes, with an empty cell
where the epoch holds no range.

Both deployments stand in one industrial hall in Torgau, Germany, so the data
cover one building and one floor. They were recorded in \DFyearSetB{}, the
NLOS deployment before the LOS deployment. The smallest rectangle that encloses the surveyed nodes in the horizontal plane
measures \qtyproduct{\DFextentLongSetB x \DFextentShortSetB}{\meter} in the LOS
deployment and \qtyproduct{\DFextentLongSetA x \DFextentShortSetA}{\meter} in
the NLOS deployment. The heights of the surveyed nodes differ by at most
\SI{\DFheightSpanCmSetB}{\centi\meter} in the LOS deployment and
\SI{\DFheightSpanCmSetA}{\centi\meter} in the NLOS deployment, so both layouts
are close to planar.

\subsection*{Recording Scenarios}
The sessions fall into three scenarios by the motion of their mobile node
(Table~\ref{tab:protocol}). In the static sessions no node moves. The NLOS
deployment holds two of them, \texttt{set\_a/s21} and \texttt{set\_a/s22}, with
all \WEnodeAnchorsStaticSetA{} nodes at surveyed positions, and in
\texttt{set\_b/s03} the tag DD98 stands still among the \WEnodeAnchorsSetB{}
anchors of the LOS deployment. In each quasi-static session \texttt{set\_a/s01}
to \texttt{s20}, the roving tag 8B05 stands on a tripod over a surveyed marker,
P01 in \texttt{s01} to P20 in \texttt{s20} (Fig.~\ref{fig:layout_b}), while the
other \WEnodeAnchorsQsSetA{} nodes stay in place. In the walks \texttt{set\_b/s01} and
\texttt{s02}, a person carried DD98 through the LOS deployment under motion
capture (Fig.~\ref{fig:layout_a}).

In the NLOS deployment, obstructions block the line of sight between many node
pairs. DD98 is a fixed anchor there, and the logs carry no timestamps, so
\texttt{seq} alone orders the epochs. The supplementary material lists every
session.

\begin{table}[!t]
  \caption{Measured sessions, grouped by the motion of the mobile node.}
  \label{tab:protocol}
  \centering
  \footnotesize
  \setlength{\tabcolsep}{4pt}
  \begin{tabular}{@{}llrrr@{}}
    \toprule
    \shortstack[l]{\strut Session} & \shortstack[l]{\strut Deployment} & \shortstack[r]{\strut Epochs} & \shortstack[r]{\strut Directed\\\strut pairs} & \shortstack[r]{\strut Ranges per\\\strut pair (median)} \\
    \midrule
    \multicolumn{5}{@{}l}{Static sessions} \\
    \texttt{set\_a/s21} & NLOS & 3640 & 130 & 3033 \\
    \texttt{set\_a/s22} & NLOS & 7626 & 126 & 6696 \\
    \texttt{set\_b/s03} & LOS & 2080 & 110 & 1783.5 \\
    \midrule
    \multicolumn{5}{@{}l}{Quasi-static sessions, rover on verified markers} \\
    \texttt{set\_a/s01}--\texttt{s20} & NLOS & 169--241 & 118--124 & 130--200 \\
    \midrule
    \multicolumn{5}{@{}l}{Walks with motion-capture truth} \\
    \texttt{set\_b/s01} & LOS & 3589 & 110 & 3145 \\
    \texttt{set\_b/s02} & LOS & 3626 & 110 & 3192.5 \\
    \bottomrule
  \end{tabular}
\end{table}

\subsection*{Ground Truth}
A Leica TS16 total station~\cite{leica_TS16} surveyed every fixed node of both
deployments, the markers and reference points of the NLOS deployment, and the
control points of the LOS deployment. Both deployments lie in local Cartesian
frames. The axes of the LOS deployment point along grid east, grid north, and
up, and its node file gives the georeference of the survey, while the NLOS
deployment lies in a campaign-local frame of the instrument. The node
files give every surveyed position a conservative prior uncertainty of
\SI{\DFsigmaSurveyCm}{\centi\meter}.

In the quasi-static sessions, the truth of 8B05 is the surveyed marker of each
session, raised by the tag height. The survey holds no position of the tag in
these sessions, so a fit to the ranges between the tag and the anchors gives
this height, and a check of the same ranges verifies each marker. The
supplementary material describes the fit and the check, which use no output of
a calibration method.

A Vicon motion-capture system~\cite{vicon_system} tracked DD98 in the LOS
deployment as one rigid object, with its markers on the tag, whose antenna
offset was not recorded. The poses lie in the Vicon frame, which one rigid
transform maps into the survey frame. The total station surveyed the
origin of the Vicon frame and a point on its positive $y$-axis, which fix the
yaw and the translation of the transform without any UWB range. Both walks store this
registration in their \texttt{session.yaml}, and the static
\texttt{set\_b/s03} has none.

The ranges and the poses carry the clocks of two computers. One offset per
session, \DFclockOffsetSpanSetB\,\si{\second} with an assumed uncertainty of
\SI{\WEgtClockSigma}{\second}, moves the ranges onto the clock of the poses,
which the files call the Vicon clock. \texttt{session.yaml} stores the offset,
and the supplementary material derives it.

The truth of the walks is sparse in time. The Vicon frame counter advances at
\WEgtViconRateSpan\,\si{\hertz}, but the congested logging chain stored the
poses at a session mean of \DFviconMeanRateSpan\,\si{\hertz} in bursts, and no
native captures were archived.

\subsection*{Processing Pipeline}
\begin{figure}[!htb]
  \centering
  \includegraphics{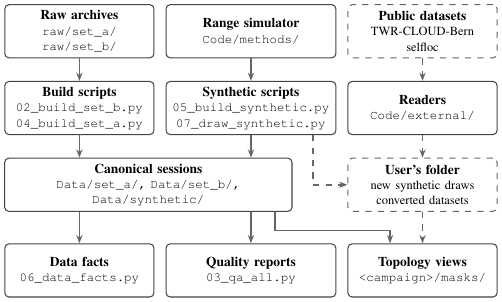}
  \caption{Processing pipeline from the three sources of the testing ground to
    the sessions and what derives from them. Dashed boxes and arrows mark data
    that the Zenodo record does not hold.}
  \label{fig:pipeline}
\end{figure}

The processing pipeline of Fig.~\ref{fig:pipeline} turns the raw files into
canonical sessions. One build script per deployment writes the node file from
the survey, parses the backup logs, and drops the failed exchanges, which the
logs record as negative or implausibly large ranges. These make up
\SI{\DFinvalidPctSetB}{\percent} of all ranges in the LOS deployment and
\SI{\DFinvalidPctSetA}{\percent} in the NLOS deployment. Every other range stays
in the data, NLOS-biased ranges included. Both build scripts also attach the
truth of their deployment, as described above.

\subsection*{Field Labels}
The file \texttt{los\_nlos\_field\_labels.csv} of each deployment gives the LOS
condition of every pair of static nodes, as judged by eye in the field.
The labels of the NLOS deployment come from the analysis code of an earlier
study~\cite{jung_uncertainty_propagation_2025}. They grade its
\DFlabelPairsSetA{} pairs as LOS, obstructed LOS (OLOS), or NLOS,
\DFlabelMixedSetA{} of them with two values, and describe the static positions
of \texttt{set\_a/s21} and \texttt{s22}. Those of the LOS deployment come from a
hand-written matrix of the campaign and cover \DFlabelPairsSetB{} pairs,
\DFlabelMixedSetB{} of them with two values. The labels record a judgment made
in the field.
Fig.~\ref{fig:layout} draws the pairs labeled NLOS, and not those with two
values. In the LOS deployment, every pair labeled NLOS includes 5BB6, 4AA4, or
D5B4, three nodes near one corner of the layout.

\subsection*{Range Simulator}
The range simulator draws synthetic ranges on a surveyed geometry with the
seeded port of the range-error model~\cite{schwarzbach_statistical_2022}. Each
range carries hardware noise, a log-normal multipath excess that becomes more
likely with the distance $d$ between its two nodes, or, with a fixed
probability, an outlier drawn
uniformly below the maximum range, and on request a range drops out with a
probability that grows with the distance. Every setting is an option of the
draw tool (Table~\ref{tab:simulator}). The noise and the multipath take the
values that the model fitted in a bus depot, whose distances also set the
maximum range, and the outlier probability, which the model leaves free, takes
that of the code that the simulator ports. These defaults are
examples. Another site calls for other values, as in a study of Bayesian
auto-positioning, which tuned the maximum range and the outlier probability per
scenario~\cite{jung_auto-positioning_2022}.

\begin{table}[!t]
  \caption{Settings of the range simulator, with their defaults and the options
    of the draw tool, which take probabilities as fractions.}
  \label{tab:simulator}
  \centering
  \footnotesize
  \setlength{\tabcolsep}{4pt}
  \begin{tabular}{@{}>{\raggedright\arraybackslash}p{0.42\columnwidth}%
      >{\raggedright\arraybackslash}p{0.22\columnwidth}%
      >{\raggedright\arraybackslash}p{0.28\columnwidth}@{}}
    \toprule
    Setting & Default & Option \\
    \midrule
    Hardware noise $\sigma_\mathrm{hw}$ & \SI{\WEsynSigmaHw}{\meter}
      & \path{--sigma-hw} \\
    Multipath excess, $R_\mathrm{mp}$ and $\sigma_\mathrm{mp}$ on the log scale
      & \DFsynMpMean{} and \DFsynMpSigma{} & \path{--mp-mean}, \path{--mp-sigma} \\
    Multipath where a uniform draw exceeds
      $b_\mathrm{mp} - s_\mathrm{mp}\,d/d_\mathrm{max}$
      & as in~\cite{schwarzbach_statistical_2022}
      & \path{--mp-base}, \path{--mp-slope} \\
    Outlier probability $p_\mathrm{out}$ & \SI{\WEsynPoutPct}{\percent}
      & \path{--p-out} \\
    Maximum range $d_\mathrm{max}$ & \SI{\DFsynDmax}{\meter} & \path{--d-max} \\
    Range bias of \texttt{biased} & \SI{\DFsynBias}{\meter} & \path{--bias} \\
    Dropouts & off & \path{--dropouts} \\
    Slant ranges & off & \path{--ranges-3d} \\
    \bottomrule
  \end{tabular}
\end{table}

The simulator combines these settings into \DFsynConditions{} preset
conditions. The condition \texttt{noise\_free} returns the generating
distances. The conditions \texttt{gaussian} and \texttt{biased} add hardware
noise to every range, the latter on top of a bias, the median anchor-pair
residual of the LOS deployment (VALIDATION AND QUALITY). The
condition \texttt{nlos\_per\_range} is the published model. The condition
\texttt{nlos}, an extension beyond it, draws the multipath excess once per pair
and adds hardware noise and outliers.

The synthetic writer \path{05_build_synthetic.py} writes the \DFsynSessions{}
frozen draws of the Zenodo record on the \DFsynAnchors{} surveyed anchors of
\texttt{set\_b/s03}, one noise-free draw and \WEsynDraws{} draws of each noisy
condition, each with both directions of all \DFsynPairs{} anchor pairs in
\WEsynEpochs{} epochs and without dropouts. Draw $k$ uses seed $k$, from which
the simulator spawns one random stream per error component. Each session stores
its condition, its seed, and the hashes of its ranges and of the simulator,
under a name that marks it as synthetic (\texttt{syn\_<condition>\_<NN>}). The draw tool \path{07_draw_synthetic.py}
draws further scenes on the surveyed geometry of any released session into a
folder outside the Zenodo record, with the anchors alone, with a tag at the
verified point of a quasi-static session, or with a tag that walks a fixed
path.

On a view whose ranges fix the geometry, such as the full anchor mesh, a method
whose range model matches noise-free ranges, which are horizontal distances
unless a scene draws slant ranges, recovers the surveyed geometry up to the
transform of the evaluation convention. The error that remains is the error
floor of the method, which separates its own faults from the effects of the
data, and the noisy conditions add known error components to the same
geometry. VALIDATION AND QUALITY compares the residuals of each condition with
those of the measured deployments.

\subsection*{Readers of Public Datasets}
The readers convert the department and reception testbeds of
selfloc~\cite{corbalan_self-localization_2023,corbalan_selfloc_2023} and a
sample of each measured directed pair of the evaluation scan of
TWR-CLOUD-Bern~\cite{xenakis_cooperative_2026,xenakis_twr-cloud-bern_2026} into
the format of UWB-ASC without redistribution. Both datasets hold static nodes
only. The user fetches the files that the fetch list \path{sources.yaml} names.
It gives the SHA-256 hash of every file that the readers parse, which pins the
version of the Bern record and the commit of the selfloc repository. The Bern
reader converts the released times of
flight to ranges, which carry no timestamps. The selfloc reader converts the
two testbeds whose ranges carry timestamps. Both rank each pair's samples as
epochs, which mark no common instant. Every node becomes an anchor at its released position, from the
laser-scan registration of Bern or the deployment files of selfloc, with a fixed
converter uncertainty that describes neither source. Nodes outside the largest
connected component of the measured pairs are dropped, and converted data keep
their source's terms.

A converted campaign carries the views \texttt{fully\_meshed} and
\texttt{multi\_hop}, and \texttt{multi\_hop} takes a gauge triple of its
measured pairs as its prior subset. The converted Bern scan is a partial mesh,
in which only part of the unordered node pairs carry ranges. A further dataset
needs a reader that returns its nodes and directed ranges, a call in the
converter, and an entry in the fetch list.

\section*{VALIDATION AND QUALITY}
Every measured session carries a quality report, \path{qa/report.md}, with its
pair availability, its residuals against the survey, and for a mobile node with
truth the coverage of its poses or the verification of its marker, and the
supplementary material summarizes these reports. The analyses below recompute such measures across the
sessions from the released files and run no estimator. Each states its
criterion.

\subsection*{Ranges Against the Survey}
\begin{figure}[!t]
  \centering
  \includegraphics{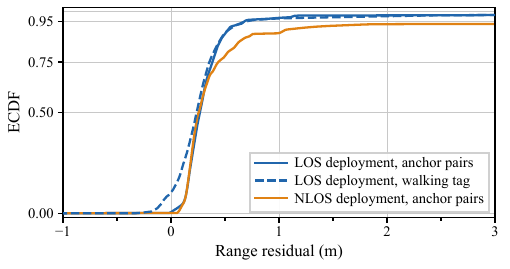}
  \caption{Empirical cumulative distribution functions of the
    residuals~\eqref{eq:residual}. Horizontal lines mark the median, P75, and
    P95. Curves ending below one continue past the abscissa, where the NLOS
    curve passes P95.}
  \label{fig:residuals}
\end{figure}

A range is compared with the distance between the true positions of its two
nodes. For the range $\tilde{d}_{ij}$ between the nodes $i$ and $j$ at the
positions $\mathbf{p}_i$ and $\mathbf{p}_j$, the residual is
\begin{equation}
  e_{ij} = \tilde{d}_{ij} - \lVert \mathbf{p}_i - \mathbf{p}_j \rVert .
  \label{eq:residual}
\end{equation}
For two anchors the positions come from the survey, and $\tilde{d}_{ij}$ is the
range of the pair in one epoch, the mean of both directions where both exist.
Fig.~\ref{fig:residuals} shows the distribution of \eqref{eq:residual} over the
epochs of the anchor pairs, in the sessions \texttt{set\_b/s01} to \texttt{s03}
of the LOS deployment and in the static sessions \texttt{set\_a/s21} and
\texttt{s22} of the NLOS deployment. The residual of a pair is the median of
\eqref{eq:residual} over its epochs in a session, and its median over the
sessions and then over the pairs is \SI{\WEresAaMedSetB}{\meter} in the LOS
deployment and \SI{\WEresAaMedSetA}{\meter} in the NLOS deployment. This median
counts every pair once, while the curves of Fig.~\ref{fig:residuals} weight
each pair by its epochs. Both values lie near the common offset of two-way
ranging that the antenna delays of the modules
cause~\cite{piavanini_calibration_2022,yudanto_device-level_2023}.

The criterion is that every pair reports its median and a robust spread, and
that each pair biased by more than a meter is named, here or in the
supplementary material. In the LOS deployment the only such pair, 5BB6 to D38F,
exceeds its surveyed distance by \DFresObstructedSpanSetB\,\si{\meter} over the
sessions. In the NLOS deployment, \DFresAaBiasedPairsSetA{} anchor pairs are
biased by \SI{\DFresAaBiasThresholdM}{\meter} or more in the median, and further
pairs by more than a meter. All of them are labeled NLOS, yet many pairs
labeled NLOS stay within a meter of their surveyed distance.

For the walking tag of the LOS deployment, $\mathbf{p}_i$ in
\eqref{eq:residual} is its registered pose matched to the epoch, and its ranges
are averaged in the same way. The median of these residuals, taken per session and
anchor and then over these, is \SI{\WEresMaMedSetB}{\meter}, close to that of
the anchor pairs. Since the registration uses surveyed points and no UWB range,
the ranges of the tag test it independently, and this agreement supports it.
For the NLOS deployment this comparison carries no evidence,
because the tag height and a common range bias are fitted to the same ranges,
and the survey checks the height instead.

\begin{figure*}[!t]
  \centering
  \subfloat[LOS deployment, \texttt{set\_b/s01}--\texttt{s03}\label{fig:availability_a}]{%
    \includegraphics{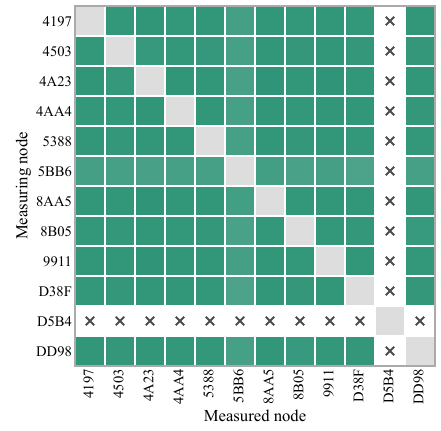}}\hfil
  \subfloat[NLOS deployment, static \texttt{set\_a/s21} and \texttt{s22}\label{fig:availability_b}]{%
    \includegraphics{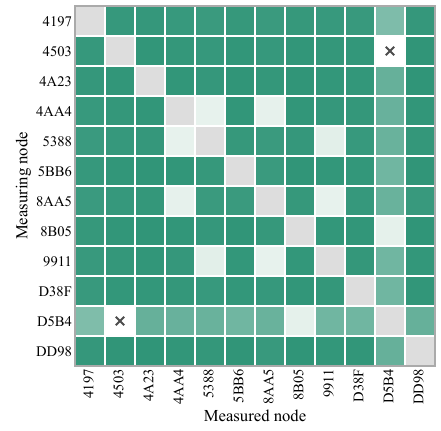}}\\[1pt]
  \includegraphics{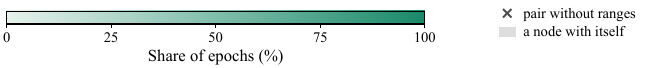}
  \caption{Availability of the directed pairs, pooled over the sessions of each
    panel.}
  \label{fig:availability}
\end{figure*}

\subsection*{Pair Availability}
Fig.~\ref{fig:availability} gives, for each cell of the logged distance matrix,
the share of the epochs in which the directed pair ranges. The criterion is that
the supplementary material names every pair without ranges, and that each group
of such pairs is given its reason or the statement that no file records one. In
the LOS deployment every directed pair among the ranging nodes carries ranges in
every session, and D5B4 never ranges. No campaign file records why, and the
field labels mark several of its pairs LOS, which speaks against an
obstruction. In the NLOS deployment, every anchor pair that ranges rarely or
never carries the field label NLOS. The roving tag 8B05 lacks one or two
anchors at several markers, some in one direction only, and no file records a
reason for these.

\subsection*{Quality of the Ground Truth}
The criterion of this analysis is that every step from the survey to the poses
states its error, its check, or that it was not recorded, and the supplementary material lists each step
with its value and its source. The nominal accuracy of one survey point comes
from the specification of the total station~\cite{leica_TS16}, apart from the
conservative prior of the node files, and it sets the uncertainty of the
registration. The motion capture's own tracking error was not measured in this
setup.

Two rules of the data-facts script match the poses to the epochs and split them
into bursts. A pose counts as the truth of a ranging epoch when it lies within
\SI{\DFmatchTolS}{\second}
of the earliest range of the epoch, less than the period of the epochs, and a
step of more than \SI{\DFmaxGapS}{\second} between poses, far longer than their
spacing within a burst, ends a burst. Under these rules the truth covers about
a fifth of the epochs of each walk. The static
\texttt{s03}, whose tag never moves, keeps a similar share, and the Vicon frame
counter advances at its full rate across every gap, which places the cause of
the gaps in the logging chain. The registration and the clock offset each add
an error at the centimeter level to the truth, the former for nearly all
matched poses and the latter at the median speed of the tag.

In the quasi-static sessions, the fitted tag height agrees with the height
above P01 that the survey gives for 8B05 in the static sessions. A position fix
from the ranges puts the tag near its marker in every session but \texttt{s05}
and \texttt{s12}, and in these two the median residual at the marker lies
within the span of the others, so every session carries verified truth.

\subsection*{Synthetic Draws Against Their Model}
The residual of a synthetic range follows \eqref{eq:residual} for every directed
range, with the horizontal distance, as the simulator generates it. The criterion is that each condition reproduces its model, within
tolerances far above the sampling error. The supplementary
material checks the noise, the bias, the excess per pair, and the tail share of
the noisy conditions against their tolerances, and every condition meets the
criterion, while the noise-free draw equals its generating distances.
Against the measured residuals, the defaults spread wider, with the noise of
the bus depot, and reach farther into the tail, with the outlier probability of
the ported code, so a user who wants draws closer to this hall sets the noise,
the outlier probability, and the maximum range accordingly.

\begin{table*}[!t]
  \caption{Files of the Zenodo record. An archive counts its members, a readme and a license among them.}
  \label{tab:files}
  \centering
  \footnotesize
  \setlength{\tabcolsep}{4pt}
  \begin{tabular}{@{}>{\raggedright\arraybackslash}p{0.37\textwidth}rrrr>{\raggedright\arraybackslash}p{0.32\textwidth}@{}}
    \toprule
    Files & Count & Size & Rows & Columns & Content \\
    \midrule
    \multicolumn{6}{@{}l}{Repository snapshot} \\
    \texttt{README.md}, \texttt{CHANGELOG.md}, \texttt{CITATION.cff}, \texttt{.zenodo.json}, \texttt{LICENSE-*} & 6 & \SI{54}{kB} & -- & -- & documentation, metadata, licenses \\
    \texttt{requirements.txt}, \texttt{PROVENANCE.yaml}, \texttt{SHA256SUMS}, \texttt{.gitattributes}, \texttt{.gitignore} & 5 & \SI{45}{kB} & -- & -- & pinned versions, source commits, checksums, checkout and ignore rules \\
    \texttt{docs/FAIR.md} & 1 & \SI{10}{kB} & -- & -- & how the release meets FAIR and best practices \\
    \texttt{Data/README.md}, \texttt{Data/set\_*/README.md} & 3 & \SI{33}{kB} & -- & -- & data dictionary, properties and limits \\
    \texttt{Data/*/campaign.yaml}, \texttt{Data/*/nodes.yaml} & 6 & \SI{16}{kB} & -- & -- & metadata, raw-file manifest, nodes, frames \\
    \texttt{Data/*/masks/*.yaml} & 12 & \SI{6}{kB} & -- & -- & the four topology views \\
    \texttt{Data/set\_*/los\_nlos\_field\_labels.csv} & 2 & \SI{3}{kB} & \num{121} & 4--5 & field LOS/NLOS labels \\
    \texttt{Data/set\_a/truth\_adjudication.csv} & 1 & \SI{2}{kB} & \num{20} & 14 & truth adjudication of the quasi-static sessions \\
    \texttt{Data/set\_a/s*/ranges.parquet} & 22 & \SI{13.8}{MB} & \num{1493969} & 6 & directed ranges, \texttt{set\_a} \\
    \texttt{Data/set\_a/s*/groundtruth.parquet} & 20 & \SI{107}{kB} & \num{20} & 10 & marker truth of the quasi-static sessions \\
    \texttt{Data/set\_b/s*/ranges.parquet} & 3 & \SI{8.3}{MB} & \num{877823} & 6 & directed ranges, \texttt{set\_b} \\
    \texttt{Data/set\_b/s*/groundtruth.parquet} & 3 & \SI{1.3}{MB} & \num{16150} & 10 & logged Vicon poses \\
    \texttt{Data/synthetic/syn\_*/ranges.parquet} & 81 & \SI{15.0}{MB} & \num{1458000} & 6 & SYNTHETIC: seeded range draws \\
    \texttt{Data/*/*/session.yaml} & 106 & \SI{110}{kB} & -- & -- & sources, clock, counts, transform, draw \\
    \texttt{Data/set\_*/s*/qa/*} & 103 & \SI{3.7}{MB} & -- & -- & quality report, residual table, figures \\
    \texttt{Code/datasets/radiocal/} & 26 & \SI{195}{kB} & -- & -- & loader, masks, the numbered scripts 01 to 07 \\
    \texttt{Code/methods/} & 2 & \SI{19}{kB} & -- & -- & modules of the range simulator \\
    \texttt{Code/external/} & 7 & \SI{33}{kB} & -- & -- & readers, converter, fetch list \\
    \texttt{Code/descriptor/} & 1 & \SI{42}{kB} & -- & -- & the script that draws Figs.~\ref{fig:views}, \ref{fig:layout}, \ref{fig:residuals}, \ref{fig:availability}, and~S1 \\
    \texttt{tests/} & 2 & \SI{18}{kB} & -- & -- & smoke test, rebuild check \\
    \midrule
    \multicolumn{6}{@{}l}{Raw part, beside the snapshot} \\
    \texttt{raw\_set\_a.zip} & 25 & \SI{9.7}{MB} & -- & -- & backup logs, survey file \\
    \texttt{raw\_set\_b.zip} & 9 & \SI{9.0}{MB} & -- & -- & backup logs, UWB and Vicon databases, survey \\
    \texttt{RAW\_MANIFEST.md} & 1 & \SI{6}{kB} & -- & -- & SHA-256 of every raw file and original database \\
    \bottomrule
  \end{tabular}
\end{table*}

\section*{RECORDS AND STORAGE}
UWB-ASC is archived as one Zenodo record~\cite{schwarzbach_uwbasc_2026} with a
version DOI and a concept DOI that resolves to the latest version. It holds a
snapshot of the GitHub repository and, beside it, two archives with the raw
files of each deployment and the manifest \texttt{RAW\_MANIFEST.md}, which lists
the SHA-256 hash of every raw file. Table~\ref{tab:files} lists every file of
the Zenodo record with its count and size, and for tables with their rows and
columns.
The data and the documentation carry the CC~BY~4.0 license and the code the MIT
license. A major version of the Zenodo record changes
canonical values or columns, a minor version adds sessions or files, and a patch
changes documentation and metadata only. All versions hang under the concept
DOI, and \path{CHANGELOG.md} lists the changes of each. No floor plan ships, so
the surveyed nodes and control points are the only geometry of the site.

These choices follow the best practices that BACKGROUND cites. The
documentation also gives the site, the hardware, the motion, and the ground
truth of each campaign, and it states what was not recorded. The loader reads
the open formats, and the readers apply both to existing datasets. The scripts
that derive the canonical files and the synthetic draws are part of the record,
and the record and this article carry identifiers of their own.
\path{docs/FAIR.md} maps these practices and the FAIR
principles~\cite{wilkinson_fair_2016} to the files that meet them, and it names
the practices that the campaigns could not follow, such as photographs of the
setup.

\subsection*{Layout and Identifiers}
The canonical data lie under \texttt{Data/}, in one folder per campaign
(\texttt{set\_a}, \texttt{set\_b}, and \texttt{synthetic}) with one subfolder
per session. A session is addressed as \texttt{<set>/<session>}, for example
\texttt{set\_b/s01}, because a bare session id occurs in both deployments.
Nodes are named by their hexadecimal ids, and views by their masks. The data
dictionary \texttt{Data/README.md} defines every column and field with
its unit and type, and the README of each deployment states its properties and
limits.

\subsection*{Files and Columns}
\texttt{ranges.parquet} holds the directed ranges of a session, with empty
times in the NLOS deployment. It keeps the direction of the logged distance
matrix, so each range is one measurement of node \texttt{to\_id} by node
\texttt{from\_id}. \texttt{groundtruth.parquet} holds the truth of the mobile
node. In the LOS deployment its poses lie in the Vicon frame, and the
transform of the session maps them into the survey frame. In the NLOS deployment
it holds one row per quasi-static session in the survey frame, the marker raised
by the tag height. \texttt{nodes.yaml} gives the frame, the nodes, and the
control points of a campaign, and Table~\ref{tab:files} names what the other
YAML files hold.

\subsection*{Relations and Formats}
The files relate through their identifiers and hashes. A session belongs to the
campaign of its folder, and every range, pose, and field label joins a node of
the \texttt{nodes.yaml} of that campaign, which the smoke test checks for every
session. Each canonical file traces back to its raw files through the
\texttt{sources} entries of \texttt{session.yaml} and \texttt{campaign.yaml}.
Views relate to the sessions through the masks only, which every campaign
carries in \texttt{masks/} and the loader ships as package data. Measurements
are columnar Parquet files~\cite{apache_parquet} and metadata are YAML, both
readable without the released code. \texttt{SHA256SUMS} lists the hash of every
file of the GitHub repository.

\section*{INSIGHTS AND NOTES}
The motion-capture truth of the walks has gaps, because the logging
chain dropped poses, and no native captures exist to fill the gaps (VALIDATION
AND QUALITY). Trajectory metrics therefore belong at the covered epochs, since
an interpolation across a gap would create poses that the system never logged.
The ranges reach the Vicon clock through an estimated offset, whose effect
VALIDATION AND QUALITY gives. The NLOS deployment has no timestamps, which
limits only a method that models the time between epochs, since its nodes stand
still within each session and the logs keep the order of the epochs.

Some nodes and pairs need care. D5B4 is surveyed in both
deployments but ranges only in the NLOS deployment. The pair 5BB6 to D38F of the
LOS deployment is obstructed and biased by meters in every session, and the
NLOS deployment holds several such pairs. These ranges stay in the data, so a
robust method meets them as they occurred. Ranges are directed, and the two
directions of a pair can differ, so an undirected method must symmetrize them.
The field labels are a reference for detecting NLOS links, while the residuals
against the survey remain the measured evidence.

The static sessions of both deployments suit the estimation of per-pair range
biases against the survey, though some NLOS biases vary, as in
\texttt{set\_a/s13} that of 9911 to D5B4. The range simulator gives controls with known errors
on the surveyed geometry, with or without a tag and with settings tuned to the
site under study, and the error floor of a method on noise-free scenes is a
first check of its implementation. The views let a study vary the topology on
the same sessions.

The scope of the testing ground is limited. It covers one hall and one hardware
family whose physical-layer settings were not recorded, its anchors stand at
nearly one height, and it holds no inertial measurements or channel impulse
responses. Uses that the author could not test include three-dimensional
calibration, methods that fuse inertial data or channel features, and the
transfer to other chip generations.

\section*{SOURCE CODE AND SCRIPTS}
The GitHub repository of UWB-ASC is published at
\url{https://github.com/PaulSchwarzbach/UWB-ASC-dataset}, and the Zenodo record
archives each of its versions~\cite{schwarzbach_uwbasc_2026}. Its README gives
the install, the commands in their order with their runtimes, and the
tolerances of a rebuild in other environments, so this article does not repeat
them.

The package \texttt{radiocal} loads a session, applies the mask of a view,
symmetrizes the ranges, and iterates over the epochs, and it carries the masks
as package data. Its frame module aligns an estimate to the survey and builds
the frame of any gauge triple. Its scripts form the processing pipeline of
Fig.~\ref{fig:pipeline}, after a check of the raw files of the LOS deployment
that writes nothing. Its data-facts script \path{06_data_facts.py} computes
every number that this article and its supplementary material print about the
data, and the residuals and the availability that the figures show. The
exceptions are the numbers of Table~\ref{tab:files}, which come from the files
themselves, and the quality reports, which the supplementary material
transcribes. A figure script draws Figs.~\ref{fig:views}, \ref{fig:layout},
\ref{fig:residuals}, and~\ref{fig:availability} and the figure of the
supplementary material from the released files and the output of the
data-facts script. Every script that writes takes
an output folder. Without it, the build scripts, the quality script, and the
synthetic writer overwrite the released files, while the draw tool refuses any
folder inside \path{Data/}. Two modules hold the range simulator, one for the
error model and one for the scenes, and both synthetic scripts import them. The
readers come with their converter and their fetch list, and the README asks
users of the readers to cite each dataset by its record and its paper.

In the environment pinned by \path{requirements.txt}, every canonical file that
the build scripts rebuild from the raw archives equals the released one, and so
does every quality figure, as \path{tests/check_rebuild.py} confirms. The smoke test \path{tests/test_smoke.py} checks an install in under
a minute. It matches all \DFsynSessions{} synthetic sessions against the hashes
of their ranges and redraws them with their seeds, both through the synthetic
writer and through the static scene of the draw tool, and given the folder of
converted campaigns, it also checks their nodes and their two views. The code
builds on NumPy~\cite{harris_array_2020}, pandas~\cite{mckinney_data_2010} with
PyArrow~\cite{apache_arrow} for Parquet~\cite{apache_parquet},
PyYAML~\cite{simonov_pyyaml_2024}, SciPy~\cite{virtanen_scipy_2020}, and
Matplotlib~\cite{hunter_matplotlib_2007}, as \path{requirements.txt} pins them.

\section*{ETHICS STATEMENT}
Not applicable, since no human participants were studied.

\section*{ACKNOWLEDGEMENTS AND INTERESTS}
P.~Schwarzbach curated and processed the data of both campaigns, developed the
software and the release, and wrote the article. The author thanks Fynn Hansen,
Albrecht Michler, and Andrea Jung for their support with the measurement
campaigns, Deveritec GmbH for the firmware that enabled the fully meshed
measurements, and MRK Management Consultants GmbH for access to the hall.

This work was done independently of any grants or awards. The author has
declared no conflicts of interest.

The author used Anthropic's Claude models through the Claude Code command-line
interface. The model versions were not pinned. The author set and drafted the
content of all sections, of the abstract, and of the supplementary material,
and the models formulated and edited the text from it. For the release, they helped
convert the recordings and two public datasets into the common
format and port the range-error model to seeded code. They also helped develop
the dataset tooling, the synthetic writer, the draw tool, the data-facts
script, the release scripts and tests, and the scripts behind the tables and
figures, and they transcribed the field labels. The author reviewed and revised
the text, the code, and the transcriptions and takes full responsibility for
this article.

\bibliographystyle{IEEEtran}
\bibliography{refs}

\end{document}


\title{Supplementary Material for ``Descriptor: UWB Anchor Self-Calibration
  Dataset (UWB-ASC)''}
\author{Paul~Schwarzbach}
\maketitle

\section*{Overview}
This material lists the measured sessions one by one, the quality report of
each session, the directed pairs without ranges, the residual and the
availability of every node pair, the field labels against the residuals, and
the quantiles of the measured and the synthetic residuals. It checks the frozen
synthetic draws against their model, and it follows the ground truth of both
deployments step by step, with the rule that verifies the markers of the
quasi-static sessions and the derivation of the clock offset of the walks.
Every value in its tables comes from the files that the data-facts script of
the release writes (\path{Code/datasets/radiocal/scripts/06_data_facts.py}), or
from the quality reports of the release. The names of the deployments,
sessions, and nodes and every definition follow the article, except the rule
that verifies the markers, which only this material states in full.

\section*{Sessions}
Table~\ref{tab:s_sessions} extends Table~\ref*{A-tab:protocol} of the article to
every measured session, with its time window, its duration, its epoch rate, and
the truth of its mobile node, from \path{sessions.csv}.

\begin{table*}[!htbp]
  \caption{Measured sessions. A dash marks a value that a session lacks.}
  \label{tab:s_sessions}
  \centering
  \scriptsize
  \setlength{\tabcolsep}{3pt}
  \begin{tabular}{@{}ll>{\raggedright\arraybackslash}p{2.3cm}lrrrrrrrrrr@{}}
    \toprule
    \shortstack[l]{\strut Session} & \shortstack[l]{\strut Motion} & \shortstack[l]{\strut Truth of the\\\strut mobile node} & \shortstack[l]{\strut Start and end\\\strut (UTC)} & \shortstack[r]{\strut Duration\\\strut (s)} & \shortstack[r]{\strut Epochs} & \shortstack[r]{\strut Epoch\\\strut rate (Hz)} & \shortstack[r]{\strut Directed\\\strut ranges} & \shortstack[r]{\strut Invalid\\\strut dropped} & \shortstack[r]{\strut Ranging\\\strut nodes} & \shortstack[r]{\strut Directed\\\strut pairs} & \shortstack[r]{\strut Ranges per\\\strut pair (median)} & \shortstack[r]{\strut Truth\\\strut poses} & \shortstack[r]{\strut Vicon\\\strut rate (Hz)} \\
    \midrule
    \multicolumn{14}{@{}l}{LOS deployment (\texttt{set\_b})} \\
    \texttt{s01} & dynamic & motion capture & 13:22:10--13:26:55 & 284.6 & 3589 & 12.61 & 339258 & 229 & 11 & 110 & 3145 & 6296 & 23.7 \\
    \texttt{s02} & dynamic & motion capture & 13:35:36--13:40:24 & 288.3 & 3626 & 12.58 & 347142 & 169 & 11 & 110 & 3192.5 & 6767 & 24.1 \\
    \texttt{s03} & static & motion capture (static, unregistered) & 13:42:54--13:45:42 & 167.9 & 2080 & 12.39 & 191423 & 185 & 11 & 110 & 1783.5 & 3087 & 23.5 \\
    \midrule
    \multicolumn{14}{@{}l}{NLOS deployment (\texttt{set\_a})} \\
    \texttt{s01} & quasi-static & marker, verified & -- & -- & 225 & -- & 21713 & 13 & 12 & 124 & 182 & 1 & -- \\
    \texttt{s02} & quasi-static & marker, verified & -- & -- & 221 & -- & 21474 & 13 & 12 & 122 & 180 & 1 & -- \\
    \texttt{s03} & quasi-static & marker, verified & -- & -- & 219 & -- & 21050 & 17 & 12 & 121 & 182 & 1 & -- \\
    \texttt{s04} & quasi-static & marker, verified & -- & -- & 170 & -- & 16092 & 19 & 12 & 122 & 137.5 & 1 & -- \\
    \texttt{s05} & quasi-static & marker, verified & -- & -- & 209 & -- & 20846 & 7 & 12 & 124 & 176.5 & 1 & -- \\
    \texttt{s06} & quasi-static & marker, verified & -- & -- & 227 & -- & 21586 & 19 & 12 & 124 & 186 & 1 & -- \\
    \texttt{s07} & quasi-static & marker, verified & -- & -- & 203 & -- & 19837 & 34 & 12 & 122 & 168 & 1 & -- \\
    \texttt{s08} & quasi-static & marker, verified & -- & -- & 189 & -- & 17860 & 16 & 12 & 122 & 156 & 1 & -- \\
    \texttt{s09} & quasi-static & marker, verified & -- & -- & 241 & -- & 23558 & 25 & 12 & 123 & 200 & 1 & -- \\
    \texttt{s10} & quasi-static & marker, verified & -- & -- & 221 & -- & 20407 & 14 & 12 & 118 & 181 & 1 & -- \\
    \texttt{s11} & quasi-static & marker, verified & -- & -- & 180 & -- & 15016 & 16 & 12 & 118 & 134.5 & 1 & -- \\
    \texttt{s12} & quasi-static & marker, verified & -- & -- & 203 & -- & 18596 & 13 & 12 & 121 & 164 & 1 & -- \\
    \texttt{s13} & quasi-static & marker, verified & -- & -- & 206 & -- & 19894 & 11 & 12 & 124 & 171 & 1 & -- \\
    \texttt{s14} & quasi-static & marker, verified & -- & -- & 174 & -- & 16322 & 18 & 12 & 122 & 140 & 1 & -- \\
    \texttt{s15} & quasi-static & marker, verified & -- & -- & 179 & -- & 16542 & 18 & 12 & 124 & 146.5 & 1 & -- \\
    \texttt{s16} & quasi-static & marker, verified & -- & -- & 169 & -- & 14384 & 6 & 12 & 122 & 130 & 1 & -- \\
    \texttt{s17} & quasi-static & marker, verified & -- & -- & 173 & -- & 16230 & 4 & 12 & 124 & 142 & 1 & -- \\
    \texttt{s18} & quasi-static & marker, verified & -- & -- & 189 & -- & 17512 & 18 & 12 & 122 & 156 & 1 & -- \\
    \texttt{s19} & quasi-static & marker, verified & -- & -- & 185 & -- & 17156 & 18 & 12 & 124 & 149.5 & 1 & -- \\
    \texttt{s20} & quasi-static & marker, verified & -- & -- & 213 & -- & 19878 & 20 & 12 & 123 & 171 & 1 & -- \\
    \texttt{s21} & static & -- & -- & -- & 3640 & -- & 351263 & 326 & 12 & 130 & 3033 & 0 & -- \\
    \texttt{s22} & static & -- & -- & -- & 7626 & -- & 766753 & 747 & 12 & 126 & 6696 & 0 & -- \\
    \bottomrule
  \end{tabular}
\end{table*}

\section*{Quality Reports}
Table~\ref{tab:s_qa} transcribes the quality report \path{qa/report.md} of each
session, which \path{03_qa_all.py} writes, as the report states it. The residual
columns give the mean and the standard deviation over the anchor pairs of the
median residual of each pair in the session, and its largest value with the
pair. The last column gives the gaps and the flagged speeds of the
motion-capture poses in the LOS deployment, and the verification of the marker
in the quasi-static sessions of the NLOS deployment, there with the median over
the directed pairs of the tag of the standard deviation of their ranges.

\begin{table*}[!htbp]
  \caption{Quality report of each measured session.}
  \label{tab:s_qa}
  \centering
  \scriptsize
  \setlength{\tabcolsep}{3pt}
  \begin{tabular}{@{}lrlrrrl>{\raggedright\arraybackslash}p{4.3cm}@{}}
    \toprule
    \shortstack[l]{\strut Session} & \shortstack[r]{\strut Directed pairs\\\strut observed} & \shortstack[l]{\strut Weakest directed pair,\\\strut share of epochs (\%)} & \shortstack[r]{\strut Anchor\\\strut pairs} & \shortstack[r]{\strut Residual\\\strut mean (m)} & \shortstack[r]{\strut Residual\\\strut std (m)} & \shortstack[l]{\strut Largest residual\\\strut (m) and its pair} & \shortstack[l]{\strut Truth of the mobile node} \\
    \midrule
    \multicolumn{8}{@{}l}{LOS deployment (\texttt{set\_b})} \\
    \texttt{s01} & 110 of 110 & D38F\,$\to$\,5BB6, 75 & 45 & +0.379 & 0.574 & +3.98, 5BB6--D38F & 26 gaps over 1\,s, the longest 36.06\,s, 12 flagged speeds \\
    \texttt{s02} & 110 of 110 & D38F\,$\to$\,5BB6, 79 & 45 & +0.381 & 0.583 & +4.05, 5BB6--D38F & 24 gaps over 1\,s, the longest 36.78\,s, 13 flagged speeds \\
    \texttt{s03} & 110 of 110 & D38F\,$\to$\,5BB6, 69 & 45 & +0.419 & 0.840 & +5.82, 5BB6--D38F & 17 gaps over 1\,s, the longest 9.14\,s, 0 flagged speeds \\
    \midrule
    \multicolumn{8}{@{}l}{NLOS deployment (\texttt{set\_a})} \\
    \texttt{s01} & 124 of 132 & 4503\,$\to$\,D5B4, 0 & 52 & +1.566 & 4.341 & +23.77, 9911--D5B4 & verified: screen 0.306\,m, range test +0.251\,m, tag ranges 0.03\,m \\
    \texttt{s02} & 122 of 132 & 4503\,$\to$\,D5B4, 0 & 51 & +1.312 & 3.873 & +23.86, 9911--D5B4 & verified: screen 0.219\,m, range test +0.205\,m, tag ranges 0.03\,m \\
    \texttt{s03} & 121 of 132 & 4503\,$\to$\,D5B4, 0 & 51 & +1.338 & 3.906 & +23.72, 9911--D5B4 & verified: screen 0.237\,m, range test +0.228\,m, tag ranges 0.07\,m \\
    \texttt{s04} & 122 of 132 & 4503\,$\to$\,D5B4, 0 & 50 & +1.290 & 3.880 & +23.72, 9911--D5B4 & verified: screen 0.436\,m, range test +0.217\,m, tag ranges 0.04\,m \\
    \texttt{s05} & 124 of 132 & 4503\,$\to$\,D5B4, 0 & 51 & +1.274 & 3.823 & +23.70, 9911--D5B4 & verified: screen 13.79\,m, range test +0.241\,m, tag ranges 0.03\,m \\
    \texttt{s06} & 124 of 132 & 4503\,$\to$\,D5B4, 0 & 51 & +1.358 & 3.967 & +24.23, 9911--D5B4 & verified: screen 0.309\,m, range test +0.176\,m, tag ranges 0.02\,m \\
    \texttt{s07} & 122 of 132 & 4503\,$\to$\,D5B4, 0 & 50 & +1.343 & 3.915 & +23.46, 9911--D5B4 & verified: screen 0.092\,m, range test +0.170\,m, tag ranges 0.04\,m \\
    \texttt{s08} & 122 of 132 & 4503\,$\to$\,D5B4, 0 & 50 & +1.325 & 3.901 & +23.62, 9911--D5B4 & verified: screen 0.195\,m, range test +0.158\,m, tag ranges 0.03\,m \\
    \texttt{s09} & 123 of 132 & 4503\,$\to$\,D5B4, 0 & 51 & +1.304 & 3.876 & +23.93, 9911--D5B4 & verified: screen 0.241\,m, range test +0.208\,m, tag ranges 0.05\,m \\
    \texttt{s10} & 118 of 132 & 4503\,$\to$\,D5B4, 0 & 50 & +1.307 & 3.890 & +23.73, 9911--D5B4 & verified: screen 0.35\,m, range test +0.231\,m, tag ranges 0.06\,m \\
    \texttt{s11} & 118 of 132 & 4503\,$\to$\,D5B4, 0 & 50 & +1.311 & 3.900 & +23.83, 9911--D5B4 & verified: screen 0.47\,m, range test +0.248\,m, tag ranges 1.33\,m \\
    \texttt{s12} & 121 of 132 & 4503\,$\to$\,D5B4, 0 & 51 & +1.302 & 3.915 & +24.24, 9911--D5B4 & verified: screen 9.93\,m, range test +0.208\,m, tag ranges 3.27\,m \\
    \texttt{s13} & 124 of 132 & 4503\,$\to$\,D5B4, 0 & 51 & +1.047 & 2.441 & +14.14, 5BB6--8AA5 & verified: screen 0.16\,m, range test +0.250\,m, tag ranges 0.02\,m \\
    \texttt{s14} & 122 of 132 & 4503\,$\to$\,D5B4, 0 & 50 & +1.311 & 3.915 & +23.98, 9911--D5B4 & verified: screen 0.24\,m, range test +0.221\,m, tag ranges 0.02\,m \\
    \texttt{s15} & 124 of 132 & 4503\,$\to$\,D5B4, 0 & 51 & +1.284 & 3.856 & +23.91, 9911--D5B4 & verified: screen 0.354\,m, range test +0.200\,m, tag ranges 0.03\,m \\
    \texttt{s16} & 122 of 132 & 4503\,$\to$\,D5B4, 0 & 51 & +1.284 & 3.889 & +24.23, 9911--D5B4 & verified: screen 0.389\,m, range test +0.207\,m, tag ranges 0.03\,m \\
    \texttt{s17} & 124 of 132 & 4503\,$\to$\,D5B4, 0 & 51 & +1.298 & 3.834 & +23.60, 9911--D5B4 & verified: screen 0.245\,m, range test +0.179\,m, tag ranges 0.08\,m \\
    \texttt{s18} & 122 of 132 & 4503\,$\to$\,D5B4, 0 & 51 & +1.298 & 3.897 & +24.19, 9911--D5B4 & verified: screen 0.348\,m, range test +0.226\,m, tag ranges 0.26\,m \\
    \texttt{s19} & 124 of 132 & 4503\,$\to$\,D5B4, 0 & 51 & +1.319 & 3.917 & +24.25, 9911--D5B4 & verified: screen 0.258\,m, range test +0.200\,m, tag ranges 0.03\,m \\
    \texttt{s20} & 123 of 132 & 4503\,$\to$\,D5B4, 0 & 51 & +1.287 & 3.847 & +23.74, 9911--D5B4 & verified: screen 0.36\,m, range test +0.200\,m, tag ranges 0.05\,m \\
    \texttt{s21} & 130 of 132 & 4503\,$\to$\,D5B4, 0 & 65 & +2.263 & 5.910 & +30.30, 8AA5--9911 & -- \\
    \texttt{s22} & 126 of 132 & 4503\,$\to$\,D5B4, 0 & 63 & +1.592 & 4.463 & +23.91, 9911--D5B4 & -- \\
    \bottomrule
  \end{tabular}
\end{table*}

\section*{Pairs Without Ranges}
Table~\ref{tab:s_missing} names the directed pairs among the ranging nodes of
each session that carry no range, from the column
\path{missing_directed_pairs} of \path{sessions.csv}. The section VALIDATION
AND QUALITY of the article names their groups, each with its reason where a
file records one, and Fig.~\ref*{A-fig:availability} there shows the availability of every
directed pair in the sessions of the LOS deployment and in the static sessions
of the NLOS deployment.

\begin{table*}[!htbp]
  \caption{Directed pairs among the ranging nodes of each measured session that carry no range.}
  \label{tab:s_missing}
  \centering
  \scriptsize
  \setlength{\tabcolsep}{3pt}
  \begin{tabular}{@{}l>{\raggedright\arraybackslash}p{15.2cm}@{}}
    \toprule
    \shortstack[l]{\strut Session} & \shortstack[l]{\strut Directed pairs without ranges} \\
    \midrule
    \multicolumn{2}{@{}l}{LOS deployment (\texttt{set\_b})} \\
    \texttt{s01} & none \\
    \texttt{s02} & none \\
    \texttt{s03} & none \\
    \midrule
    \multicolumn{2}{@{}l}{NLOS deployment (\texttt{set\_a})} \\
    \texttt{s01} & 4503\,$\to$\,D5B4, 4AA4\,$\to$\,8AA5, 5388\,$\to$\,9911, 8AA5\,$\to$\,9911, 9911\,$\to$\,5388, 9911\,$\to$\,8AA5, D5B4\,$\to$\,4503, D5B4\,$\to$\,8B05 \\
    \texttt{s02} & 4503\,$\to$\,D5B4, 4AA4\,$\to$\,8AA5, 5388\,$\to$\,9911, 8AA5\,$\to$\,4AA4, 8AA5\,$\to$\,9911, 8B05\,$\to$\,D5B4, 9911\,$\to$\,5388, 9911\,$\to$\,8AA5, D5B4\,$\to$\,4503, D5B4\,$\to$\,8B05 \\
    \texttt{s03} & 4503\,$\to$\,D5B4, 4AA4\,$\to$\,5388, 4AA4\,$\to$\,8AA5, 5388\,$\to$\,9911, 8AA5\,$\to$\,4AA4, 8AA5\,$\to$\,9911, 8B05\,$\to$\,D5B4, 9911\,$\to$\,5388, 9911\,$\to$\,8AA5, D5B4\,$\to$\,4503, D5B4\,$\to$\,8B05 \\
    \texttt{s04} & 4503\,$\to$\,D5B4, 4AA4\,$\to$\,5388, 4AA4\,$\to$\,8AA5, 5388\,$\to$\,4AA4, 5388\,$\to$\,9911, 8AA5\,$\to$\,4AA4, 8AA5\,$\to$\,9911, 9911\,$\to$\,5388, 9911\,$\to$\,8AA5, D5B4\,$\to$\,4503 \\
    \texttt{s05} & 4503\,$\to$\,D5B4, 4AA4\,$\to$\,8AA5, 5388\,$\to$\,9911, 8AA5\,$\to$\,4AA4, 8AA5\,$\to$\,9911, 9911\,$\to$\,5388, 9911\,$\to$\,8AA5, D5B4\,$\to$\,4503 \\
    \texttt{s06} & 4503\,$\to$\,D5B4, 4AA4\,$\to$\,8AA5, 5388\,$\to$\,9911, 8AA5\,$\to$\,4AA4, 8AA5\,$\to$\,9911, 9911\,$\to$\,5388, 9911\,$\to$\,8AA5, D5B4\,$\to$\,4503 \\
    \texttt{s07} & 4503\,$\to$\,D5B4, 4AA4\,$\to$\,5388, 4AA4\,$\to$\,8AA5, 5388\,$\to$\,4AA4, 5388\,$\to$\,9911, 8AA5\,$\to$\,4AA4, 8AA5\,$\to$\,9911, 9911\,$\to$\,5388, 9911\,$\to$\,8AA5, D5B4\,$\to$\,4503 \\
    \texttt{s08} & 4503\,$\to$\,D5B4, 4AA4\,$\to$\,5388, 4AA4\,$\to$\,8AA5, 5388\,$\to$\,4AA4, 5388\,$\to$\,9911, 8AA5\,$\to$\,4AA4, 8AA5\,$\to$\,9911, 9911\,$\to$\,5388, 9911\,$\to$\,8AA5, D5B4\,$\to$\,4503 \\
    \texttt{s09} & 4503\,$\to$\,D5B4, 4AA4\,$\to$\,5388, 4AA4\,$\to$\,8AA5, 5388\,$\to$\,9911, 8AA5\,$\to$\,4AA4, 8AA5\,$\to$\,9911, 9911\,$\to$\,5388, 9911\,$\to$\,8AA5, D5B4\,$\to$\,4503 \\
    \texttt{s10} & 4503\,$\to$\,D5B4, 4AA4\,$\to$\,5388, 4AA4\,$\to$\,8AA5, 4AA4\,$\to$\,8B05, 5388\,$\to$\,4AA4, 5388\,$\to$\,9911, 8AA5\,$\to$\,4AA4, 8AA5\,$\to$\,9911, 8B05\,$\to$\,4AA4, 8B05\,$\to$\,9911, 9911\,$\to$\,5388, 9911\,$\to$\,8AA5, 9911\,$\to$\,8B05, D5B4\,$\to$\,4503 \\
    \texttt{s11} & 4503\,$\to$\,D5B4, 4AA4\,$\to$\,5388, 4AA4\,$\to$\,8AA5, 4AA4\,$\to$\,8B05, 5388\,$\to$\,4AA4, 5388\,$\to$\,9911, 8AA5\,$\to$\,4AA4, 8AA5\,$\to$\,9911, 8B05\,$\to$\,4AA4, 8B05\,$\to$\,9911, 9911\,$\to$\,5388, 9911\,$\to$\,8AA5, 9911\,$\to$\,8B05, D5B4\,$\to$\,4503 \\
    \texttt{s12} & 4503\,$\to$\,D5B4, 4AA4\,$\to$\,8AA5, 5388\,$\to$\,9911, 5BB6\,$\to$\,8B05, 8AA5\,$\to$\,4AA4, 8AA5\,$\to$\,9911, 8B05\,$\to$\,9911, 9911\,$\to$\,5388, 9911\,$\to$\,8AA5, 9911\,$\to$\,8B05, D5B4\,$\to$\,4503 \\
    \texttt{s13} & 4503\,$\to$\,D5B4, 4AA4\,$\to$\,8AA5, 5388\,$\to$\,9911, 8AA5\,$\to$\,4AA4, 8AA5\,$\to$\,9911, 9911\,$\to$\,5388, 9911\,$\to$\,8AA5, D5B4\,$\to$\,4503 \\
    \texttt{s14} & 4503\,$\to$\,D5B4, 4AA4\,$\to$\,5388, 4AA4\,$\to$\,8AA5, 5388\,$\to$\,4AA4, 5388\,$\to$\,9911, 8AA5\,$\to$\,4AA4, 8AA5\,$\to$\,9911, 9911\,$\to$\,5388, 9911\,$\to$\,8AA5, D5B4\,$\to$\,4503 \\
    \texttt{s15} & 4503\,$\to$\,D5B4, 4AA4\,$\to$\,8AA5, 5388\,$\to$\,9911, 8AA5\,$\to$\,4AA4, 8AA5\,$\to$\,9911, 9911\,$\to$\,5388, 9911\,$\to$\,8AA5, D5B4\,$\to$\,4503 \\
    \texttt{s16} & 4503\,$\to$\,D5B4, 4AA4\,$\to$\,8AA5, 5388\,$\to$\,9911, 8AA5\,$\to$\,4AA4, 8AA5\,$\to$\,8B05, 8AA5\,$\to$\,9911, 8B05\,$\to$\,8AA5, 9911\,$\to$\,5388, 9911\,$\to$\,8AA5, D5B4\,$\to$\,4503 \\
    \texttt{s17} & 4503\,$\to$\,D5B4, 4AA4\,$\to$\,8AA5, 5388\,$\to$\,9911, 8AA5\,$\to$\,4AA4, 8AA5\,$\to$\,9911, 9911\,$\to$\,5388, 9911\,$\to$\,8AA5, D5B4\,$\to$\,4503 \\
    \texttt{s18} & 4503\,$\to$\,D5B4, 4AA4\,$\to$\,8AA5, 5388\,$\to$\,9911, 8AA5\,$\to$\,4AA4, 8AA5\,$\to$\,8B05, 8AA5\,$\to$\,9911, 8B05\,$\to$\,8AA5, 9911\,$\to$\,5388, 9911\,$\to$\,8AA5, D5B4\,$\to$\,4503 \\
    \texttt{s19} & 4503\,$\to$\,D5B4, 4AA4\,$\to$\,8AA5, 5388\,$\to$\,9911, 8AA5\,$\to$\,4AA4, 8AA5\,$\to$\,9911, 9911\,$\to$\,5388, 9911\,$\to$\,8AA5, D5B4\,$\to$\,4503 \\
    \texttt{s20} & 4503\,$\to$\,D5B4, 4AA4\,$\to$\,8AA5, 5388\,$\to$\,4AA4, 5388\,$\to$\,9911, 8AA5\,$\to$\,4AA4, 8AA5\,$\to$\,9911, 9911\,$\to$\,5388, 9911\,$\to$\,8AA5, D5B4\,$\to$\,4503 \\
    \texttt{s21} & 4503\,$\to$\,D5B4, D5B4\,$\to$\,4503 \\
    \texttt{s22} & 4503\,$\to$\,D5B4, 5388\,$\to$\,9911, 8AA5\,$\to$\,9911, 9911\,$\to$\,5388, 9911\,$\to$\,8AA5, D5B4\,$\to$\,4503 \\
    \bottomrule
  \end{tabular}
\end{table*}

\section*{Residual and Availability of Every Pair}
Table~\ref{tab:s_pairs} gives every node pair of the residual sessions with its
field label, its surveyed distance, its residual and spread, and its
availability, from \path{pairs.csv}. The residual sessions and the residual of
a pair follow the section VALIDATION AND QUALITY of the article. The
interquartile range is that of the residuals of a pair in the epochs of these
sessions, for the walking tag in its matched epochs, and the availability is
the share of the epochs with a range in either direction. The heading of each
group names the sessions of its residuals and of its ranges.

\begingroup
\scriptsize
\setlength{\tabcolsep}{3pt}
\begin{longtable}{@{}llrrrrrrrl@{}}
  \multicolumn{10}{c}{\parbox[b]{0.8\textwidth}{\centering\footnotesize\label{tab:s_pairs}TABLE~\thetable\\ \textsc{Residual and availability of every node pair in the residual sessions of each deployment, $a$ the first node of a pair and $b$ the second. A dash marks a pair without residual.}}} \\[8pt]
  \toprule
  \shortstack[l]{\strut Pair} & \shortstack[l]{\strut Field\\\strut label} & \shortstack[r]{\strut Surveyed\\\strut distance (m)} & \shortstack[r]{\strut Median\\\strut residual (m)} & \shortstack[r]{\strut Interquartile\\\strut range (m)} & \shortstack[r]{\strut Session\\\strut medians (m)} & \shortstack[r]{\strut Ranges\\\strut $a\to b$} & \shortstack[r]{\strut Ranges\\\strut $b\to a$} & \shortstack[r]{\strut Epochs with\\\strut a range (\%)} & \shortstack[l]{\strut Remark} \\
  \midrule
  \endfirsthead
  \toprule
  \shortstack[l]{\strut Pair} & \shortstack[l]{\strut Field\\\strut label} & \shortstack[r]{\strut Surveyed\\\strut distance (m)} & \shortstack[r]{\strut Median\\\strut residual (m)} & \shortstack[r]{\strut Interquartile\\\strut range (m)} & \shortstack[r]{\strut Session\\\strut medians (m)} & \shortstack[r]{\strut Ranges\\\strut $a\to b$} & \shortstack[r]{\strut Ranges\\\strut $b\to a$} & \shortstack[r]{\strut Epochs with\\\strut a range (\%)} & \shortstack[l]{\strut Remark} \\
  \midrule
  \endhead
  \bottomrule
  \endlastfoot
  \multicolumn{10}{@{}l}{LOS deployment (\texttt{set\_b}), anchor pairs, residuals and ranges from \texttt{s01}, \texttt{s02}, \texttt{s03}} \\
  4197--4503 & LOS & 14.60 & 0.22 & 0.021 & 0.22 to 0.23 & 8108 & 8108 & 87.23 &  \\
  4197--4A23 & LOS & 11.30 & 0.53 & 0.080 & 0.53 to 0.54 & 8227 & 8227 & 88.51 &  \\
  4197--4AA4 & NLOS & 18.91 & 0.48 & 0.041 & 0.47 to 0.48 & 8107 & 8097 & 87.26 &  \\
  4197--5388 & LOS & 16.44 & 0.17 & 0.016 & 0.17 & 8142 & 8139 & 87.60 &  \\
  4197--5BB6 & LOS & 22.13 & 0.30 & 0.027 & 0.29 to 0.30 & 7337 & 7337 & 78.93 &  \\
  4197--8AA5 & LOS & 15.80 & 0.29 & 0.020 & 0.29 to 0.30 & 8193 & 8194 & 88.18 &  \\
  4197--8B05 & LOS & 8.68 & 0.20 & 0.014 & 0.20 & 8108 & 8108 & 87.23 &  \\
  4197--9911 & LOS & 4.19 & 0.20 & 0.016 & 0.20 & 8157 & 8158 & 87.78 &  \\
  4197--D38F & LOS & 11.99 & 0.27 & 0.019 & 0.27 to 0.28 & 8122 & 8122 & 87.38 &  \\
  4197--D5B4 & NLOS & 19.61 & -- & -- & -- & 0 & 0 & 0.00 & node never ranges \\
  4503--4A23 & LOS & 8.76 & 0.25 & 0.016 & 0.25 & 7969 & 7998 & 87.46 &  \\
  4503--4AA4 & LOS & 5.53 & 0.17 & 0.029 & 0.17 & 8131 & 8129 & 87.51 &  \\
  4503--5388 & LOS & 3.16 & 0.02 & 0.033 & 0.01 to 0.02 & 8059 & 8061 & 86.82 &  \\
  4503--5BB6 & NLOS & 9.47 & 0.31 & 0.021 & 0.31 & 7334 & 7334 & 78.90 &  \\
  4503--8AA5 & LOS & 5.83 & 0.17 & 0.021 & 0.17 & 8149 & 8148 & 87.71 &  \\
  4503--8B05 & LOS & 11.06 & 0.16 & 0.019 & 0.15 to 0.16 & 8096 & 8096 & 87.10 &  \\
  4503--9911 & LOS & 11.93 & 0.27 & 0.021 & 0.27 & 8142 & 8138 & 87.61 &  \\
  4503--D38F & LOS & 11.55 & 0.15 & 0.028 & 0.15 to 0.16 & 8095 & 8095 & 87.09 &  \\
  4503--D5B4 & LOS & 9.48 & -- & -- & -- & 0 & 0 & 0.00 & node never ranges \\
  4A23--4AA4 & LOS & 14.27 & 0.67 & 0.029 & 0.67 & 8223 & 8223 & 88.47 &  \\
  4A23--5388 & LOS & 11.89 & 0.23 & 0.015 & 0.23 & 8200 & 8198 & 88.22 &  \\
  4A23--5BB6 & NLOS & 18.22 & 0.46 & 0.044 & 0.46 & 7187 & 7175 & 77.63 &  \\
  4A23--8AA5 & LOS & 13.80 & 0.40 & 0.025 & 0.40 & 8267 & 8267 & 88.95 &  \\
  4A23--8B05 & LOS & 13.74 & 0.21 & 0.018 & 0.21 & 8180 & 8178 & 88.00 &  \\
  4A23--9911 & LOS & 7.16 & 0.41 & 0.028 & 0.40 to 0.41 & 8250 & 8251 & 88.77 &  \\
  4A23--D38F & LOS & 16.06 & 0.30 & 0.030 & 0.30 & 8185 & 8185 & 88.06 &  \\
  4A23--D5B4 & LOS & 17.90 & -- & -- & -- & 0 & 0 & 0.00 & node never ranges \\
  4AA4--5388 & NLOS & 2.58 & 0.38 & 0.039 & 0.38 & 8157 & 8160 & 87.80 &  \\
  4AA4--5BB6 & LOS & 3.95 & 0.15 & 0.019 & 0.15 & 7333 & 7331 & 78.91 &  \\
  4AA4--8AA5 & LOS & 4.26 & 0.24 & 0.020 & 0.24 & 8189 & 8187 & 88.11 &  \\
  4AA4--8B05 & LOS/NLOS & 13.08 & 0.29 & 0.025 & 0.29 & 8120 & 8121 & 87.37 &  \\
  4AA4--9911 & NLOS & 16.91 & 0.99 & 0.276 & 0.98 to 0.99 & 8177 & 8177 & 87.98 &  \\
  4AA4--D38F & LOS/NLOS & 12.08 & 0.33 & 0.037 & 0.33 & 8135 & 8135 & 87.52 &  \\
  4AA4--D5B4 & LOS & 5.26 & -- & -- & -- & 0 & 0 & 0.00 & node never ranges \\
  5388--5BB6 & NLOS & 6.41 & 0.34 & 0.031 & 0.34 to 0.35 & 7320 & 7323 & 78.78 &  \\
  5388--8AA5 & LOS & 3.47 & 0.19 & 0.036 & 0.19 & 8180 & 8178 & 88.02 &  \\
  5388--8B05 & LOS & 11.21 & 0.13 & 0.021 & 0.13 & 8091 & 8091 & 87.06 &  \\
  5388--9911 & LOS & 14.33 & 0.23 & 0.023 & 0.23 to 0.24 & 8164 & 8166 & 87.85 &  \\
  5388--D38F & LOS & 10.77 & 0.13 & 0.040 & 0.12 to 0.16 & 8127 & 8127 & 87.43 &  \\
  5388--D5B4 & LOS & 6.46 & -- & -- & -- & 0 & 0 & 0.00 & node never ranges \\
  5BB6--8AA5 & LOS & 6.41 & 0.38 & 0.030 & 0.37 to 0.38 & 7389 & 7390 & 79.51 &  \\
  5BB6--8B05 & NLOS & 15.31 & 0.41 & 0.033 & 0.40 to 0.41 & 7150 & 7152 & 77.17 &  \\
  5BB6--9911 & LOS & 20.50 & 0.44 & 0.025 & 0.44 to 0.45 & 7364 & 7364 & 79.24 &  \\
  5BB6--D38F & NLOS & 13.53 & 4.05 & 0.404 & 3.98 to 5.82 & 7001 & 7001 & 75.32 &  \\
  5BB6--D5B4 & LOS & 4.15 & -- & -- & -- & 0 & 0 & 0.00 & node never ranges \\
  8AA5--8B05 & LOS & 9.02 & 0.15 & 0.026 & 0.15 & 8130 & 8128 & 87.50 &  \\
  8AA5--9911 & LOS & 14.56 & 0.53 & 0.024 & 0.51 to 0.53 & 8198 & 8201 & 88.25 &  \\
  8AA5--D38F & LOS/NLOS & 7.82 & 0.33 & 0.049 & 0.33 to 0.34 & 8169 & 8169 & 87.89 &  \\
  8AA5--D5B4 & NLOS & 4.22 & -- & -- & -- & 0 & 0 & 0.00 & node never ranges \\
  8B05--9911 & LOS & 9.82 & 0.19 & 0.017 & 0.19 & 8122 & 8125 & 87.41 &  \\
  8B05--D38F & LOS & 3.31 & 0.08 & 0.039 & 0.07 to 0.09 & 8081 & 8081 & 86.94 &  \\
  8B05--D5B4 & NLOS & 11.96 & -- & -- & -- & 0 & 0 & 0.00 & node never ranges \\
  9911--D38F & LOS & 12.97 & 0.32 & 0.031 & 0.31 to 0.32 & 8158 & 8158 & 87.77 &  \\
  9911--D5B4 & NLOS & 18.69 & -- & -- & -- & 0 & 0 & 0.00 & node never ranges \\
  D38F--D5B4 & NLOS & 9.74 & -- & -- & -- & 0 & 0 & 0.00 & node never ranges \\
  \midrule
  \multicolumn{10}{@{}l}{LOS deployment (\texttt{set\_b}), pairs of the walking tag, residuals from \texttt{s01}, \texttt{s02}, ranges from \texttt{s01}, \texttt{s02}, \texttt{s03}} \\
  4197--DD98 & -- & -- & 0.17 & 0.160 & 0.14 to 0.20 & 8136 & 8136 & 87.53 &  \\
  4503--DD98 & -- & -- & 0.11 & 0.259 & 0.10 to 0.13 & 8070 & 8069 & 86.85 &  \\
  4A23--DD98 & -- & -- & 0.24 & 0.132 & 0.23 to 0.26 & 8190 & 8192 & 88.14 &  \\
  4AA4--DD98 & -- & -- & 0.37 & 0.323 & 0.34 to 0.40 & 8094 & 8098 & 87.18 &  \\
  5388--DD98 & -- & -- & 0.10 & 0.260 & 0.09 to 0.10 & 8114 & 8115 & 87.33 &  \\
  5BB6--DD98 & -- & -- & 0.42 & 0.217 & 0.39 to 0.45 & 7277 & 7271 & 78.32 &  \\
  8AA5--DD98 & -- & -- & 0.24 & 0.235 & 0.18 to 0.30 & 8155 & 8155 & 87.77 &  \\
  8B05--DD98 & -- & -- & 0.20 & 0.213 & 0.20 to 0.21 & 7938 & 7932 & 85.59 &  \\
  9911--DD98 & -- & -- & 0.26 & 0.125 & 0.25 to 0.28 & 8128 & 8131 & 87.54 &  \\
  D38F--DD98 & -- & -- & 0.32 & 0.421 & 0.30 to 0.34 & 8084 & 8084 & 86.97 &  \\
  D5B4--DD98 & -- & -- & -- & -- & -- & 0 & 0 & 0.00 & node never ranges \\
  \midrule
  \multicolumn{10}{@{}l}{NLOS deployment (\texttt{set\_a}), anchor pairs, residuals and ranges from \texttt{s21}, \texttt{s22}} \\
  4197--4503 & LOS & 15.62 & 0.22 & 0.013 & 0.22 & 9671 & 9638 & 85.90 &  \\
  4197--4A23 & LOS & 11.66 & 0.40 & 0.013 & 0.40 & 9747 & 9727 & 86.53 &  \\
  4197--4AA4 & LOS & 35.85 & 0.19 & 0.020 & 0.19 to 0.20 & 9629 & 9602 & 85.56 &  \\
  4197--5388 & OLOS & 30.96 & 0.19 & 0.017 & 0.19 & 9474 & 9449 & 84.19 &  \\
  4197--5BB6 & LOS & 35.56 & 0.73 & 0.020 & 0.73 to 0.74 & 9703 & 9676 & 86.14 &  \\
  4197--8AA5 & LOS & 32.57 & 0.20 & 0.025 & 0.20 to 0.21 & 9484 & 9453 & 84.28 &  \\
  4197--8B05 & LOS & 7.77 & 0.08 & 0.018 & 0.08 to 0.09 & 9722 & 9698 & 86.34 &  \\
  4197--9911 & LOS & 39.02 & 0.31 & 0.022 & 0.31 to 0.32 & 9605 & 9583 & 85.37 &  \\
  4197--D38F & LOS & 24.97 & 0.16 & 0.021 & 0.16 & 9570 & 9570 & 84.95 &  \\
  4197--D5B4 & NLOS & 28.80 & 3.80 & 0.054 & 3.80 & 5728 & 5689 & 52.02 &  \\
  4197--DD98 & LOS & 20.54 & 0.07 & 0.021 & 0.07 & 9765 & 9733 & 86.69 &  \\
  4503--4A23 & LOS & 19.60 & 0.14 & 0.016 & 0.14 & 9932 & 9932 & 88.23 &  \\
  4503--4AA4 & OLOS & 35.93 & 0.22 & 0.026 & 0.22 & 9789 & 9796 & 87.07 &  \\
  4503--5388 & NLOS & 18.92 & 4.97 & 1.239 & 4.56 to 5.38 & 9670 & 9668 & 85.97 &  \\
  4503--5BB6 & OLOS & 38.77 & 0.32 & 0.023 & 0.31 to 0.32 & 9902 & 9903 & 87.95 &  \\
  4503--8AA5 & NLOS & 24.10 & 1.27 & 0.220 & 1.20 to 1.35 & 9662 & 9666 & 85.96 &  \\
  4503--8B05 & LOS & 7.88 & 0.15 & 0.015 & 0.15 & 9907 & 9915 & 88.07 &  \\
  4503--9911 & OLOS & 35.83 & 0.27 & 0.025 & 0.27 & 9807 & 9807 & 87.16 &  \\
  4503--D38F & OLOS & 19.91 & 0.22 & 0.028 & 0.21 to 0.23 & 9824 & 9824 & 87.20 &  \\
  4503--D5B4 & NLOS & 15.10 & -- & -- & -- & 0 & 0 & 0.00 & no ranges \\
  4503--DD98 & LOS & 25.88 & 0.16 & 0.021 & 0.16 & 9961 & 9961 & 88.49 &  \\
  4A23--4AA4 & LOS & 24.63 & 0.23 & 0.023 & 0.23 to 0.24 & 9896 & 9898 & 87.92 &  \\
  4A23--5388 & LOS & 27.40 & 0.19 & 0.020 & 0.19 & 9758 & 9754 & 86.65 &  \\
  4A23--5BB6 & LOS & 23.90 & 0.18 & 0.018 & 0.18 & 9965 & 9965 & 88.45 &  \\
  4A23--8AA5 & NLOS & 25.88 & 0.61 & 0.028 & 0.61 & 9773 & 9779 & 86.86 &  \\
  4A23--8B05 & LOS & 13.64 & 0.16 & 0.017 & 0.16 & 9985 & 9982 & 88.65 &  \\
  4A23--9911 & LOS & 28.76 & 0.46 & 0.051 & 0.46 to 0.47 & 9853 & 9859 & 87.58 &  \\
  4A23--D38F & LOS & 17.31 & 0.18 & 0.028 & 0.17 to 0.18 & 9855 & 9855 & 87.48 &  \\
  4A23--D5B4 & LOS & 27.26 & 0.28 & 0.018 & 0.27 to 0.28 & 7056 & 7057 & 62.69 &  \\
  4A23--DD98 & LOS & 8.88 & 0.13 & 0.017 & 0.12 to 0.13 & 10026 & 10026 & 88.99 &  \\
  4AA4--5388 & NLOS & 27.84 & 0.87 & 0.032 & 0.87 to 0.88 & 29 & 14 & 0.35 &  \\
  4AA4--5BB6 & LOS & 7.48 & 0.42 & 0.023 & 0.41 to 0.43 & 9866 & 9863 & 87.64 &  \\
  4AA4--8AA5 & NLOS & 19.72 & 16.40 & 0.091 & 16.39 to 16.41 & 55 & 66 & 0.77 &  \\
  4AA4--8B05 & LOS & 34.49 & 0.15 & 0.022 & 0.14 to 0.15 & 9897 & 9897 & 87.93 &  \\
  4AA4--9911 & LOS & 7.86 & 0.21 & 0.027 & 0.21 to 0.22 & 9749 & 9743 & 86.67 &  \\
  4AA4--D38F & LOS & 16.86 & 0.28 & 0.030 & 0.27 to 0.28 & 9812 & 9812 & 87.09 &  \\
  4AA4--D5B4 & NLOS & 31.79 & 1.56 & 0.322 & 1.54 to 1.58 & 6923 & 6934 & 61.69 &  \\
  4AA4--DD98 & LOS & 16.44 & 0.24 & 0.021 & 0.23 to 0.25 & 9917 & 9910 & 88.06 &  \\
  5388--5BB6 & NLOS & 33.79 & 0.59 & 0.038 & 0.58 to 0.59 & 9708 & 9705 & 86.23 &  \\
  5388--8AA5 & LOS & 8.72 & 0.18 & 0.017 & 0.18 & 9463 & 9463 & 84.16 &  \\
  5388--8B05 & NLOS & 24.16 & 0.50 & 0.201 & 0.44 to 0.55 & 9730 & 9727 & 86.45 &  \\
  5388--9911 & NLOS & 23.78 & 16.60 & 0.190 & 16.60 & 277 & 278 & 2.48 & residual from \texttt{s21} only \\
  5388--D38F & NLOS & 12.91 & 0.51 & 0.040 & 0.49 to 0.52 & 9615 & 9615 & 85.35 &  \\
  5388--D5B4 & LOS & 5.21 & 0.14 & 0.019 & 0.14 to 0.15 & 6877 & 6880 & 61.13 &  \\
  5388--DD98 & LOS & 27.72 & 0.16 & 0.017 & 0.15 to 0.16 & 9782 & 9788 & 86.89 &  \\
  5BB6--8AA5 & NLOS & 26.26 & 14.13 & 0.029 & 14.13 to 14.14 & 9505 & 9499 & 84.65 &  \\
  5BB6--8B05 & LOS & 35.83 & 0.19 & 0.027 & 0.19 to 0.20 & 9947 & 9947 & 88.30 &  \\
  5BB6--9911 & LOS & 15.28 & 0.51 & 0.021 & 0.50 to 0.51 & 9816 & 9817 & 87.22 &  \\
  5BB6--D38F & LOS & 21.62 & 0.35 & 0.019 & 0.35 to 0.36 & 9827 & 9827 & 87.23 &  \\
  5BB6--D5B4 & NLOS & 37.19 & 0.60 & 0.038 & 0.59 to 0.61 & 6476 & 6496 & 58.30 &  \\
  5BB6--DD98 & LOS & 15.02 & 0.33 & 0.014 & 0.32 to 0.33 & 9970 & 9969 & 88.50 &  \\
  8AA5--8B05 & OLOS/NLOS & 27.15 & 0.29 & 0.030 & 0.29 & 9713 & 9710 & 86.35 &  \\
  8AA5--9911 & NLOS & 15.08 & 30.30 & 16.445 & 30.30 & 3 & 3 & 0.04 & residual from \texttt{s21} only \\
  8AA5--D38F & NLOS & 8.63 & 1.03 & 0.028 & 1.03 & 9666 & 9666 & 85.80 &  \\
  8AA5--D5B4 & LOS & 13.59 & 0.31 & 0.032 & 0.31 to 0.32 & 6871 & 6872 & 61.14 &  \\
  8AA5--DD98 & NLOS & 23.42 & 0.67 & 0.027 & 0.66 to 0.67 & 9558 & 9556 & 85.15 &  \\
  8B05--9911 & LOS & 36.12 & 0.11 & 0.020 & 0.11 & 9852 & 9856 & 87.56 &  \\
  8B05--D38F & LOS & 20.70 & 0.19 & 0.025 & 0.19 to 0.20 & 9859 & 9859 & 87.51 &  \\
  8B05--D5B4 & NLOS & 21.44 & 17.22 & 3.635 & 16.94 to 17.51 & 111 & 101 & 1.42 &  \\
  8B05--DD98 & LOS & 21.49 & 0.22 & 0.017 & 0.21 to 0.22 & 10006 & 10008 & 88.85 &  \\
  9911--D38F & LOS & 15.93 & 0.29 & 0.021 & 0.29 & 9784 & 9784 & 86.85 &  \\
  9911--D5B4 & NLOS & 28.44 & 23.84 & 0.391 & 23.76 to 23.91 & 6488 & 6481 & 58.27 &  \\
  9911--DD98 & LOS & 21.83 & 0.35 & 0.021 & 0.35 to 0.36 & 9854 & 9847 & 87.52 &  \\
  D38F--D5B4 & NLOS & 15.63 & 0.41 & 0.036 & 0.40 to 0.41 & 6506 & 6506 & 57.75 &  \\
  D38F--DD98 & LOS & 15.21 & 0.27 & 0.016 & 0.27 to 0.28 & 9865 & 9865 & 87.56 &  \\
  D5B4--DD98 & LOS & 29.20 & 0.21 & 0.023 & 0.21 to 0.22 & 7047 & 7044 & 62.60 &  \\
\end{longtable}
\endgroup

\section*{Field Labels Against the Residuals}
Table~\ref{tab:s_labels} counts the labeled pairs of each field label and
gives the median and the largest of their residuals, from \path{labels.csv}.
The labels are the field judgment that the section COLLECTION METHODS AND
DESIGN of the article describes, and the residuals do not correct them.

\begin{table*}[!htbp]
  \caption{Field labels against the residuals of the labeled pairs.}
  \label{tab:s_labels}
  \centering
  \footnotesize
  \setlength{\tabcolsep}{3pt}
  \begin{tabular}{@{}lrrrrr@{}}
    \toprule
    \shortstack[l]{\strut Field label} & \shortstack[r]{\strut Pairs} & \shortstack[r]{\strut Pairs with\\\strut a residual} & \shortstack[r]{\strut Median\\\strut residual (m)} & \shortstack[r]{\strut Largest\\\strut residual (m)} & \shortstack[r]{\strut Pairs at or above\\\strut \SI{\DFresAaBiasThresholdM}{\meter}} \\
    \midrule
    \multicolumn{6}{@{}l}{LOS deployment (\texttt{set\_b})} \\
    LOS & 39 & 34 & 0.23 & 0.67 & 0 \\
    NLOS & 13 & 8 & 0.43 & 4.05 & 0 \\
    LOS/NLOS & 3 & 3 & 0.33 & 0.33 & 0 \\
    \midrule
    \multicolumn{6}{@{}l}{NLOS deployment (\texttt{set\_a})} \\
    LOS & 40 & 40 & 0.21 & 0.73 & 0 \\
    OLOS & 5 & 5 & 0.22 & 0.32 & 0 \\
    NLOS & 20 & 19 & 1.27 & 30.30 & 6 \\
    OLOS/NLOS & 1 & 1 & 0.29 & 0.29 & 0 \\
    \bottomrule
  \end{tabular}
\end{table*}

\section*{Quantiles of the Residuals}
Table~\ref{tab:s_accuracy} gives the quantiles of the measured residuals of
Fig.~\ref*{A-fig:residuals} of the article and of the residuals of the frozen
synthetic draws, from \path{residual_quantiles.csv} and
\path{synthetic_quantiles.csv}, rounded to the centimeter, among them the values
beyond the abscissa of Fig.~\ref*{A-fig:residuals}, such as the P95 of the NLOS
deployment.

\begin{table*}[!htbp]
  \caption{Quantiles of the range residuals~\textup{(\ref*{A-eq:residual})} of the article, measured as in its Fig.~\ref*{A-fig:residuals} and synthetic as in Fig.~\ref{fig:s_synthetic}, in meters.}
  \label{tab:s_accuracy}
  \centering
  \footnotesize
  \setlength{\tabcolsep}{4pt}
  \begin{tabular}{@{}lrrrrr@{}}
    \toprule
    Residuals & P5 & P25 & Median & P75 & P95 \\
    \midrule
    \multicolumn{6}{@{}l}{Measured, against the survey} \\
    LOS deployment, anchor pairs & \num{0.11} & \num{0.19} & \num{0.27} & \num{0.38} & \num{0.68} \\
    LOS deployment, walking tag & \num{-0.09} & \num{0.12} & \num{0.24} & \num{0.36} & \num{0.69} \\
    NLOS deployment, anchor pairs & \num{0.12} & \num{0.18} & \num{0.26} & \num{0.44} & \num{3.89} \\
    \midrule
    \multicolumn{6}{@{}l}{Synthetic, seeded draws per condition} \\
    \texttt{noise\_free} & \num{0.00} & \num{0.00} & \num{0.00} & \num{0.00} & \num{0.00} \\
    \texttt{gaussian} & \num{-0.54} & \num{-0.22} & \num{0.00} & \num{0.22} & \num{0.54} \\
    \texttt{biased} & \num{-0.27} & \num{0.05} & \num{0.27} & \num{0.49} & \num{0.81} \\
    \texttt{nlos\_per\_range} & \num{-0.51} & \num{-0.15} & \num{0.14} & \num{0.61} & \num{8.27} \\
    \texttt{nlos} & \num{-0.52} & \num{-0.16} & \num{0.11} & \num{0.52} & \num{7.96} \\
    \bottomrule
  \end{tabular}
\end{table*}

\section*{Synthetic Draws Against Their Model}
\begin{figure}[!htbp]
  \centering
  \includegraphics{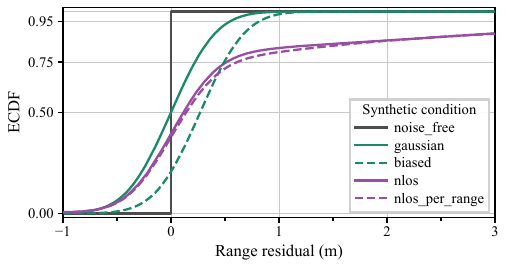}
  \caption{Empirical cumulative distribution functions of the residuals of the
    frozen synthetic draws against their generating distances, one curve per
    condition. Horizontal grid lines mark the median, P75, and P95. Curves
    that start above zero or end below one continue beyond the abscissa.}
  \label{fig:s_synthetic}
\end{figure}

Fig.~\ref{fig:s_synthetic} gives the distribution of the residuals of the frozen
synthetic draws against their generating distances, per condition and pooled
over the draws, from \path{synthetic_quantiles.csv}, on the abscissa and the
grid lines of Fig.~\ref*{A-fig:residuals} of the article. The section VALIDATION
AND QUALITY of the article states the criterion of these draws, and the
statistics below meet it.

Each noisy condition pools \num{\DFsynRangesPerCondition} ranges over its
draws, and the residual follows \textup{(\ref*{A-eq:residual})} of the article. The
\texttt{gaussian}
residuals spread with a standard deviation of \SI{\DFsynResSdGaussian}{\meter}
and the \texttt{biased} ones center on a median of
\SI{\DFsynResMedBiased}{\meter}, within \SI{\DFsynTolM}{\meter} of
$\sigma_\mathrm{hw}$ and of the bias. In every \texttt{nlos} draw, the median
residual of each pair lies within \SI{\DFsynNlosExcessDevMax}{\meter} of its
stored excess, against a tolerance of $\DFsynNlosTolSigma\,\sigma_\mathrm{hw}$.
Of the \texttt{nlos\_per\_range} residuals, \SI{\DFsynTailSharePct}{\percent}
exceed $\DFsynTailSigma\,\sigma_\mathrm{hw}$, where the published model
predicts \SI{\DFsynTailShareModelPct}{\percent} for the drawn distances, within
\DFsynTailTolPct{} percentage points.

In Table~\ref{tab:s_accuracy}, the measured residuals center near the offset of
two-way ranging, and their interquartile ranges are narrower than that of the
\texttt{gaussian} preset by more than the averaging of the two directions of a
measured range explains, since averaging narrows a Gaussian spread by at most
the square root of two. The tails of \texttt{nlos\_per\_range} and
\texttt{nlos} reach farther than that of the NLOS deployment, mostly through the
outliers drawn uniformly below the maximum range.

\clearpage
\section*{Ground Truth}
Table~\ref{tab:s_truth} follows the ground truth of both deployments from the
survey to the poses, each step with its value and its source. VALIDATION AND
QUALITY of the article states the criterion of this check and how the poses are
matched to the epochs.

In the quasi-static sessions of the NLOS deployment, the truth rests on a rule
in three steps, and \path{04_build_set_a.py} writes its statistics to
\path{truth_adjudication.csv}. First, a screen by a two-dimensional
multilateration from the median ranges of 8B05 to the static nodes passes a
session whose tag lies within \SI{\DFscreenDevM}{\meter} of its marker. Second,
a robust joint fit over the sessions that pass returns the tag height and one
common range bias from the median ranges between the tag and the anchors.
Third, in a range test, a session that the screen flags passes only if the
median residual of its LOS links, those within
\SI{\DFlosTolM}{\meter} of the fitted bias, falls within the span of the
sessions that passed the screen. The tolerance was chosen on the flagged
sessions.

In the LOS deployment, the ranges and the poses passed one MQTT broker, which
stamped each message on receipt. The difference between receipt and stamp of a
message holds the offset of its sender's clock plus a transport latency, so the
offset between the two clocks is the difference of the smallest such values of
the two streams, whose shortest latencies nearly cancel.

\begin{table*}[!htbp]
  \caption{Ground truth of both deployments, step by step, with the value and
    the source of each step.}
  \label{tab:s_truth}
  \centering
  \footnotesize
  \setlength{\tabcolsep}{4pt}
  \begin{tabular}{@{}>{\raggedright\arraybackslash}p{0.40\textwidth}%
      >{\raggedright\arraybackslash}p{0.22\textwidth}%
      >{\raggedright\arraybackslash}p{0.34\textwidth}@{}}
    \toprule
    Step & Value & Source \\
    \midrule
    \multicolumn{3}{@{}l}{Both deployments} \\
    Survey point, nominal accuracy & \SI{\DFsurveyPointSigmaMm}{\milli\meter}
      & TS16 specification, cited in the article \\
    \midrule
    \multicolumn{3}{@{}l}{LOS deployment (\texttt{set\_b})} \\
    Offset of the antenna of DD98 from the origin of the tracked object, whose
      markers sat on the tag
      & not recorded & \path{Data/set_b/README.md} \\
    Registration, yaw uncertainty from two surveyed points
      \SI{\WEgtWandBaseline}{\meter} apart
      & \SI{\WEgtYawSigma}{\degree} & \path{session.yaml} \\
    Its effect within \SI{\WEgtYawRadius}{\meter} of the origin of the Vicon
      frame, where nearly all matched poses lie
      & \SI{\WEgtYawEffect}{\meter}
      & \path{groundtruth.parquet}, \path{ranges.parquet}, \path{session.yaml} \\
    Epochs with a matched pose, walks and static \texttt{s03}
      & \WEgtCoverShareSpan\,\% and \SI{\WEgtCoverShareStatic}{\percent}
      & \path{groundtruth.parquet}, \path{ranges.parquet}, \path{session.yaml} \\
    Longest gap between poses, \texttt{s01}, \texttt{s02}, and \texttt{s03}
      & \DFgapMaxBone, \DFgapMaxBtwo, and \SI{\DFgapMaxBthree}{\second}
      & \path{groundtruth.parquet} \\
    Timing error at the median speed of the tag over
      \SI{\DFspeedHorizonS}{\second}, walks
      & \WEgtClockSpan\,\si{\meter}
      & \path{groundtruth.parquet}, \path{session.yaml} \\
    \midrule
    \multicolumn{3}{@{}l}{NLOS deployment (\texttt{set\_a})} \\
    Tag height above the marker, fitted and surveyed
      & \SI{\WEgtTagHeight}{\meter} and \SI{\WEgtTagHeightSurvey}{\meter}
      & \path{truth_adjudication.csv}, \path{nodes.yaml} \\
    Common range bias of the fit & \SI{\WEgtRangeBiasSetA}{\meter}
      & \path{truth_adjudication.csv} \\
    Screen: sessions passed, and the deviation of the flagged \texttt{s05} and
      \texttt{s12}
      & \DFscreenPassSetA, then \SI{\WEgtScreenDevAfive}{\meter} and
        \SI{\WEgtScreenDevAtwelve}{\meter}
      & \path{truth_adjudication.csv} \\
    Range test: reference span, then \texttt{s05} and \texttt{s12}
      & \WEgtRangeTestSpan\,\si{\meter}, then \SI{\WEgtRangeTestAfive}{\meter} and
        \SI{\WEgtRangeTestAtwelve}{\meter}
      & \path{truth_adjudication.csv} \\
    Sessions with verified truth & all & \path{truth_adjudication.csv} \\
    \bottomrule
  \end{tabular}
\end{table*}